\documentclass[aps,prb,amsmath,superscriptaddress,twocolumn,bibnotes,longbibliography,floatfix]{revtex4-2}

\usepackage{amsfonts}
\usepackage{graphicx}
\usepackage{color}
\usepackage{dsfont}
\usepackage{verbatim}
\usepackage{mathtools}
\usepackage[normalem]{ulem}
\usepackage{comment}
\usepackage{stackrel}
\usepackage{bm}
\usepackage{booktabs}

\usepackage[colorlinks=true, 
linkcolor=blue, 
urlcolor=blue,
citecolor=blue]{hyperref}
\usepackage{orcidlink}
\usepackage[linesnumbered]{algorithm2e}

\RestyleAlgo{ruled}
\SetKwComment{Comment}{/* }{ */}
\definecolor{myblue}{rgb}{.93, .93, 1}

\newcommand{\bsub}{\begin{subequations}}
	\newcommand{\esub}{\end{subequations}}

\newcommand{\vex}[1]{\bm{\mathrm{#1}}}

\newcommand{\tr}{\mathsf{Tr}}

\begin{document}
	
	\title{Topological flat bands, valley polarization, and interband superconductivity in magic-angle twisted bilayer graphene with proximitized spin-orbit couplings}
	
	\author{Yang-Zhi~Chou~\orcidlink{0000-0001-7955-0918}}\email{yzchou@umd.edu}
	\affiliation{Condensed Matter Theory Center and Joint Quantum Institute, Department of Physics, University of Maryland, College Park, Maryland 20742, USA}
	
	\author{Yuting~Tan~\orcidlink{0000-0002-9600-5986}}
	\email{ytan77@umd.edu}
	\affiliation{Condensed Matter Theory Center and Joint Quantum Institute, Department of Physics, University of Maryland, College Park, Maryland 20742, USA}
	
	\author{Fengcheng~Wu~\orcidlink{0000-0002-1185-0073}}
	\affiliation{School of Physics and Technology, Wuhan University, Wuhan  430072, China}
	\affiliation{Wuhan Institute of Quantum Technology, Wuhan 430206, China}
	
	\author{Sankar Das~Sarma~\orcidlink{0000-0002-0439-986X}}
	\affiliation{Condensed Matter Theory Center and Joint Quantum Institute, Department of Physics, University of Maryland, College Park, Maryland 20742, USA}
	
	\date{\today}

	\begin{abstract}
		We study theoretically the magic-angle twisted bilayer graphene with proximity-induced Ising and Rashba spin-orbit couplings on the top layer. Topological flat bands (with three distinct phases) are generically realized by the spin-orbit couplings. Using a mean field analysis, we find that (partial) valley polarization prevails for a wide range of doping, suppressing the usual superconductivity with a pairing between time-reversal partners. Remarkably, we uncover that observable unconventional intervalley interband phonon-mediated superconductivity (with the highest $T_c\approx 1.2$K) can coexist with strong valley imbalance due to the approximate Fermi surface nesting between two flat bands not related by time-reversal symmetry, and the dominant pairing is an intersublattice Ising pairing, corresponding to a mixture of $p$- and $d$-waves. In contrast, the intrasublattice Ising phonon-mediated superconductivity with $s$- and $f$-wave mixing emerges in the absence of valley imbalance. Our work reveals an unprecedented route of realizing unconventional superconductivity. 
	\end{abstract}
	
	\maketitle
	
	\textit{Introduction. ---} Graphene, with or without moir\'e, provides controllable ways to create novel quantum phases, such as correlated insulators and superconductivity (SC) \cite{CaoY2018a,CaoY2018,PolshynH2019,SharpeAL2019,JiangY2019,LuX2019,YankowitzM2019,KerelskyA2019,XieY2019,PolshynH2019,ChoiY2019,CaoY2020,SerlinM2020,AroraHS2020,WongD2020,ChoiY2021,OhM2021,LiuX2020,HaoZ2021,ZhouH2021,ZhouH2021a,ChoiY2021,ParkJM2021,ParkJM2021a,LiuX2021,ZhouH2022,KuiriM2022,ZhangY2023,HolleisL2023,SuR2023,PixleyJH2019,AndreiEY2020,BalentsL2020,AndreiEY2021}. 
	Recent experiments have incorporated spin-orbit coupling (SOC) in graphene systems induced by a proximate WSe$_2$ layer, offering a novel way to manipulate graphene-based materials \cite{IslandJO2019,AroraHS2020,ZhangY2023,HolleisL2023,SuR2023,SunL2023,XieM2023}. In particular, observable SC can be induced \cite{AroraHS2020,SuR2023} or enhanced \cite{ZhangY2023,HolleisL2023} by the proximate WSe$_2$ layer, and the interplay between SOC and SC in graphene is an active area of research \cite{ChouYZ2022,CurtisJB2023,Jimeno-PozoA2023,PantaleonPA2023,ParappurathA2023,WagnerG2023}. Meanwhile, the intrinsic SOC is believed to be crucial for the phase diagram of rhombohedral trilayer graphene \cite{ArpT2023b}. The goal of this Letter is to study the magic-angle twisted bilayer graphene (MATBG) with proximity-induced SOCs, in which the bandwidth is comparable to the induced SOC.

	\begin{figure}[t!]
		\includegraphics[width=0.45\textwidth]{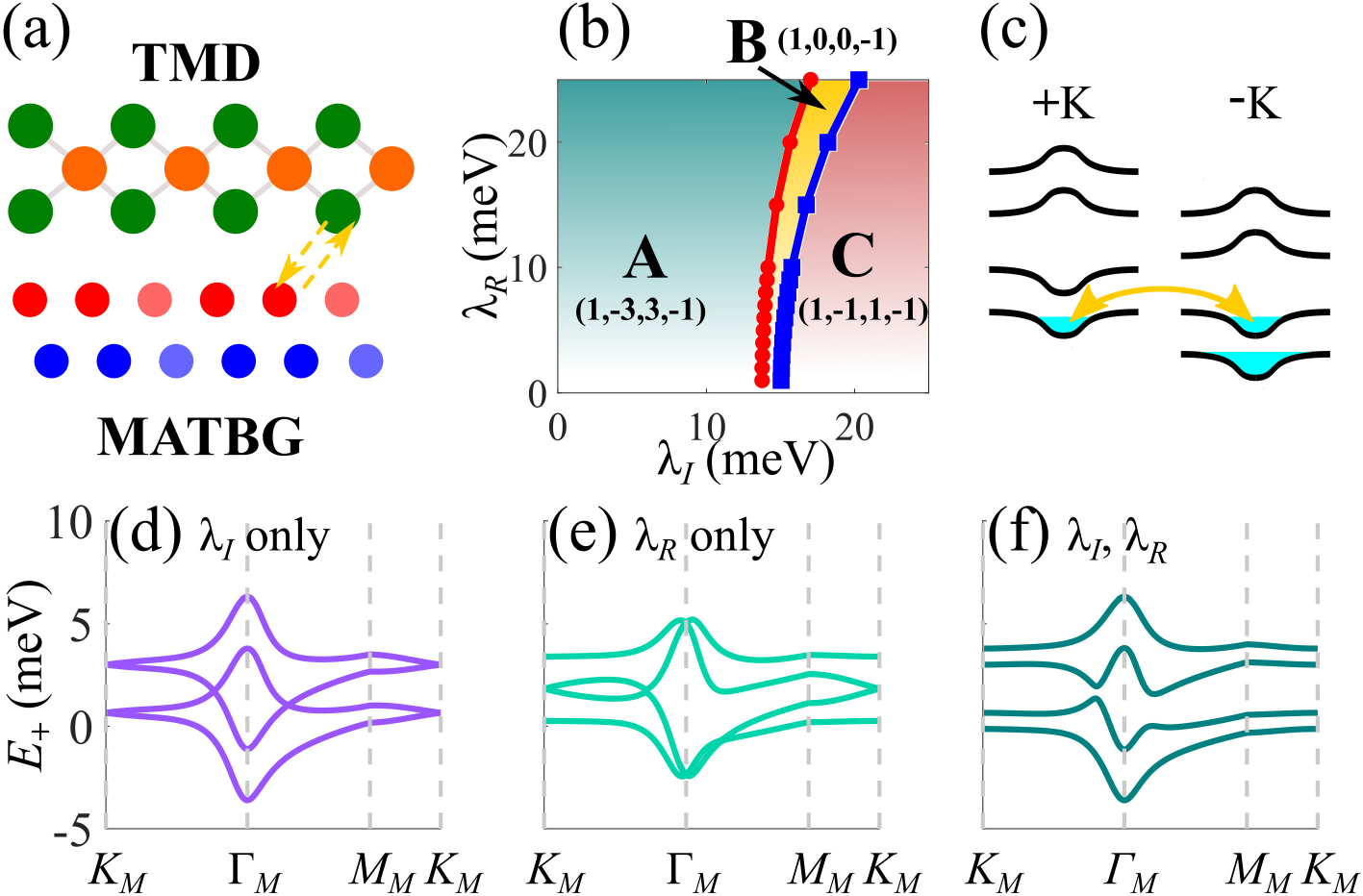}
		\caption{Setup and main results. (a) The setup of the system. The MATBG system is proximate to a substrate (e.g., WSe$_2$) that can induce SOC on the top graphene layer (red). The bottom graphene layer (blue) is unaffected. (b) The single-particle topological phase diagram characterized by the Chern numbers with $\theta=1.05^\circ$.  (c) Intervalley interband pairing in a (partially) valley-polarized state. (d)-(f) MATBG band structure of $\theta=1.05^\circ$. (d) $\lambda_I=5$meV and $\lambda_R=0$. (e) $\lambda_I=0$ and $\lambda_R=5$meV. (f) $\lambda_I=5$meV and $\lambda_R=5$meV.}
		\label{Fig:Setup}
	\end{figure}	
	
	We investigate the low-temperature phases of the MATBG with proximitized SOCs (Ising and Rashba) on the top layer, as illustrated in Fig~\ref{Fig:Setup}(a). 
	First, we reveal topological flat bands with nontrivial spin textures and construct a single-particle topological phase diagram [Fig.~\ref{Fig:Setup}(b)]. Using a mean field (MF) analysis on the projected Hamiltonian, we then identify the (partial) valley polarization (VP) as the leading instability for a wide range of fillings, twist angles, and SOC parameters. The presence of valley imbalance breaks the time-reversal symmetry and suppresses the intervalley pairings between time-reversal partners. Remarkably, we find that interband intersublattice Ising phonon-mediated SC ($p$- and $d$-wave mixing) can coexist with the strong valley imbalance [as illustrated in Fig.~\ref{Fig:Setup}(c)] due to the approximate Fermi surface nesting in the flat bands. This contrasts with the intrasublattice Ising pairing ($s$- and $f$-wave mixing) predicted by the acoustic-phonon mechanism \cite{WuF2019,ChouYZ2021,ChouYZ2022} in the absence of VP.

	\textit{Model. ---} We consider MATBG with proximity-induced Ising and Rashba SOCs on the top graphene layer. The proximity-induced SOCs can be realized by placing a transition metal dichalcogenide layer (e.g., WSe$_2$) on top of MATBG. Using the Bistritzer-MacDonald model \cite{BistritzerR2011,WuF2018}, the low-energy Hamiltonian of the $+K$ valley is expressed by
	\begin{align}\label{Eq:H_0_BM}
		\mathcal{H}_{0,+}=\left[\begin{array}{cc}
			\hat{U}_{\theta/2}\left(\hat{h}_t^{(+)}(\vex{k})+\hat{h}^{(+)}_{\text{SOC},t}\right)\hat{U}_{\theta/2}^{\dagger} & \hat{T}^{\dagger}(\vex{x})\\[2mm]
			\hat{T}(\vex{x}) & \hat{U}_{\theta/2}^{\dagger}\hat{h}_b^{(+)}(\vex{k})\hat{U}_{\theta/2}
		\end{array}\right],
	\end{align}
	where $\theta$ is the twist angle, and the subscripts $t$ and $b$ denote the top and bottom layers, respectively. In the above expression, $\hat{h}_t^{(+)}$ and $\hat{h}_b^{(+)}$ are the isolated $+K$ valley Dirac Hamiltonian of the top and the bottom layers with the expressions $\hat{h}_l^{(+)}(\vex{k})=v_F\left(\vex{k}-\vex{\kappa}_l\right)\cdot\vex{\sigma}$ for $l=t,b$.
	Here, $v_F\approx5.944\text{eV\AA}$ is the Dirac velocity of the monolayer graphene \cite{SongZ2019}, $\vex{\sigma}=(\sigma_x,\sigma_y)$, $\sigma_{\mu}$ is the $\mu$-component Pauli matrix for the sublattice, and $\vex{\kappa}_t$ ($\vex{\kappa}_b$) is the rotated $+K$ valley point of the top (bottom) layer. To encode the rotation of the spinors in the Dirac Hamiltonian, we also apply the sublattice rotation matrix $\hat{U}_{\theta/2}=e^{i(\theta/4)\sigma_z}$. The interlayer tunneling between two twisted layers induces a spatially varying potential, described by $\hat{T}(\vex{x})=\hat{t}_0+\hat{t}_{1}e^{-i\vex{b}_+\cdot\vex{x}}+\hat{t}_{-1}e^{-i\vex{b}_-\cdot\vex{x}}$, where $\hat{t}_j=w_0\sigma_0+w_1[\cos(2\pi j/3)\sigma_x+\sin(2\pi j/3)\sigma_y]$, $w_1\approx 110$meV, $w_0=0.8w_1$, $\vex{b}_{\pm}=[4\pi/(\sqrt{3}a_M)]\left(\pm1/2,\sqrt{3}/2\right)$, and $a_M$ is the moir\'e lattice constant. Finally, we consider SOC terms in the $\tau K$ valley given by \cite{AroraHS2020,NaimerT2021,NaimerT2023,WangT2020,ScammellHD2023a,ScammellHD2024}
	\begin{align}\label{Eq:h_SOC}
		\hat{h}_{\text{SOC},t}^{(\tau)}=&\frac{\lambda_I}{2}\tau s_z+\frac{\lambda_R}{2}\left(\tau\sigma_xs_y-\sigma_ys_x\right),
	\end{align}
	where $\lambda_I$ ($\lambda_R$) denotes the Ising (Rashba) SOC strength and $s_{\mu}$ is the $\mu$-component Pauli matrix for the spins. 
	
	The presence of SOCs alters the underlying symmetry of MATBG. The overall system (including both $+K$ and $-K$ valleys) obeys the spinful time-reversal symmetry, $\mathcal{T}_s=i\tau_xs_y\mathcal{K}$, where $\tau_x$ is the $x$-component Pauli matrix for the valley and $\mathcal{K}$ is the conjugation operator. Thus, we expect the moir\'e bands of the two valleys satisfy $\mathcal{E}_{+,b}(\vex{k})=\mathcal{E}_{-,b}(-\vex{k})$ and $\Psi_{-,b,-\vex{k}}=i\tau_xs_y\Psi_{+,b,\vex{k}}^*$ for the energies and wavefunctions of the $b$th band, respectively. Our model satisfies the spinful three-fold rotational symmetry about the $z$ axis.

	Before proceeding, we discuss several issues concerning the proximity-induced SOCs. In Eq.~(\ref{Eq:h_SOC}), we have simplified the expression for the $\lambda_R$ term \cite{NaimerT2021,NaimerT2023,ZollnerK2023a,FrankT2024}. The induced SOC strengths depend on the relative angle between the SOC layer and the top graphene layer \cite{LiY2019,NaimerT2021,DavidA2019,ChouYZ2022,ZollnerK2023a}, and $\lambda_I$ and $\lambda_R$ can be as large as a few meV, comparable to the bandwidth of MATBG. Here, we treat $\lambda_I$ and $\lambda_R$ as free parameters and study a range of the twist angle.

	\begin{figure}[t!]
		\includegraphics[width=0.425\textwidth]{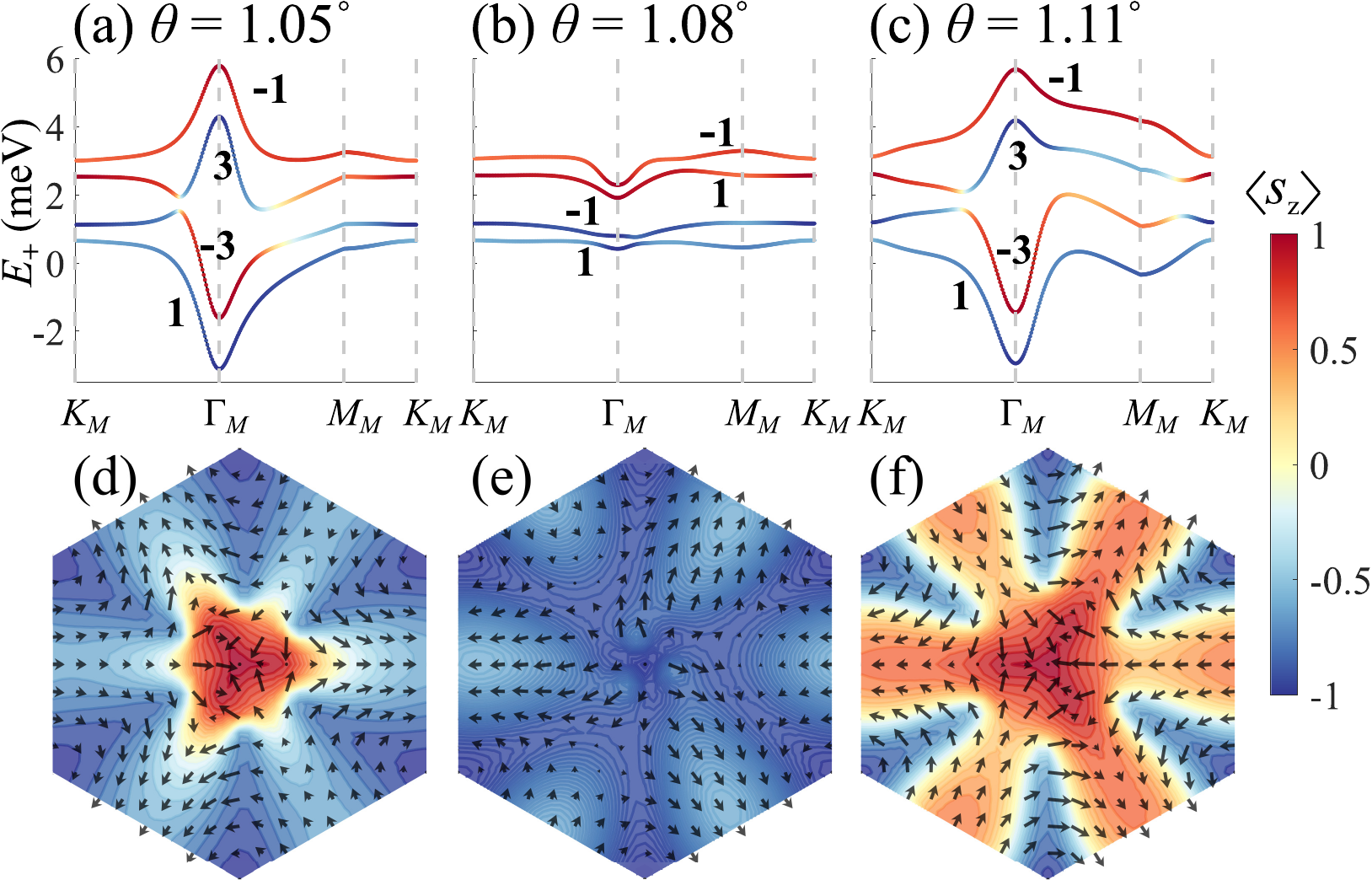}
		\caption{Dispersion and spin texture in the $+K$ valley. $\lambda_I=\lambda_R=3$meV are used for all the plots. (a), (b), and (c) are the energy bands along high symmetry points for $\theta=1.05^\circ$, $\theta=1.08^\circ$, and $\theta=1.11^\circ$, respectively. We also mark the Chern numbers for each band. (d), (e), and (f) show the spin texture of the second band (from the bottom) in the first moir\'e Brillouin zone associated with (a), (b), and (c). The color indicates the expectation of the $z$-component spin; the arrows show the spin component in the $xy$ plane.
		}
		\label{Fig:Band_angle}
	\end{figure}

	\textit{Band structure and topology. ---} We diagonalize the single-particle Hamiltonian $\mathcal{H}_{0,+}$ [Eq.~(\ref{Eq:H_0_BM})] in the momentum space. Without SOC, the smallest bandwidth ($\approx 1.1$meV) is found around $\theta=1.08^{\circ}$, which we define as the magic angle. For $\lambda_I\neq 0$ and $\lambda_R=0$ [Fig.~\ref{Fig:Setup}(d)], the band structure is shifted following the valley-spin Zeeman field in Eq.~(\ref{Eq:h_SOC}). For $\lambda_I= 0$ and $\lambda_R\neq0$ [Fig.~\ref{Fig:Setup}(e)], a single Dirac band touching at the $K_M$ ($K_M'$) point exists.
	As shown in Fig.~\ref{Fig:Setup}(f), the combination of both Ising and Rashba SOCs reconstructs the moir\'e bands significantly, generically resulting in four spin-split bands around charge neutrality.
	
	%To characterize the topological properties of the moir\'e bands induced by SOCs, 
	We extract the Chern number of each band by numerically computing the Wilson loops in a momentum-space rhombus grid \cite{ChouYZ2020}. The Chern number of the $b$th band in $\pm K$ valley is labeled by $\mathcal{C}_{\pm, b}$. Due to the time-reversal symmetry, $\mathcal{C}_{-, b}=-\mathcal{C}_{+, b}$. In Fig.~\ref{Fig:Setup}(b), we identify three distinct phases, described by $(\mathcal{C}_{+,1},\mathcal{C}_{+,2},\mathcal{C}_{+,3},\mathcal{C}_{+,4})=(1,-3,3,-1)$, $(1,0,0,-1)$, and $(1,-1,1,-1)$, coined phases A, B, and C, respectively. The phases A \cite{WangT2020,LinJX2022,BhowmikS2023} and C \cite{WangT2020} were previously reported, and we point out that phase B can also be realized with Rashba and Ising SOCs. By varying $\lambda_R$ and $\lambda_I$, $\mathcal{C}_{+, 1}$ and $\mathcal{C}_{+, 4}$ remain unchanged (as long as $\lambda_I>0$ and $\lambda_R\neq 0$), and the topological transitions happen in the middle two bands associated with the emergence of Dirac nodes. The Chern numbers change sign with a negative $\lambda_I$, while the sign of $\lambda_R$ does not affect the Chern numbers. At the transition between A and B phases, three Dirac nodes develop near the $\Gamma_M$ point in the middle two bands, consistent with the Chern numbers changing by 3 passing the transition and topological systems with three-fold rotational symmetry \cite{FangC2012}. At the phase boundary between B and C, a Dirac node develops at the $\Gamma_M$ point in the middle two bands, corresponding to the Chern number changing by 1. The phase diagram of the $\theta=1.05^\circ$ case is shown in Fig.~\ref{Fig:Setup}(b), and similar phase diagrams can be constructed for different choices of $\theta$. Notably, the transition between B and C roughly traces $\lambda_I=2w_{0,\theta}$, where $w_{0,\theta}$ is the $\Gamma_M$ point energy difference between conduction and valence bands for twist angle $\theta$ in the absence of SOC.

	We also investigate the twist-angle dependence of the single-particle properties with fixed SOC strengths. In Fig.~\ref{Fig:Band_angle}, the band structures and spin textures with $\lambda_I=\lambda_R=3$meV are plotted for three different values of $\theta$. The $\theta=1.08^\circ$ case [Fig.~\ref{Fig:Band_angle}(b)] shows four nearly flat bands which belong to the phase C, and the lower (higher) two bands have $\langle s_z\rangle<0$ ($\langle s_z\rangle>0$), suggesting small hybridization between two spin states. Meanwhile, the $\theta=1.05^\circ$ [Fig.~\ref{Fig:Band_angle}(a)] and $\theta=1.11^\circ$ [Fig.~\ref{Fig:Band_angle}(c)] cases, belonging to phase A, show larger bandwidths and significant spin mixing. In Figs.~\ref{Fig:Band_angle}(d)-(f), we plot the spin textures of the second bands corresponding to Figs.~\ref{Fig:Band_angle}(a)-(c), respectively. Remarkably, $\theta=1.05^\circ$ [Fig.~\ref{Fig:Band_angle}(d)] and $\theta=1.11^\circ$ [Fig.~\ref{Fig:Band_angle}(f)] cases manifest skyrmion-like textures: the $\langle s_z\rangle$ near the $\Gamma_M$ and the $K_M$ points have different signs. A detailed study of the momentum-space spin configurations will be reported elsewhere.

	\begin{figure}[t!]
		\includegraphics[width=0.4\textwidth]{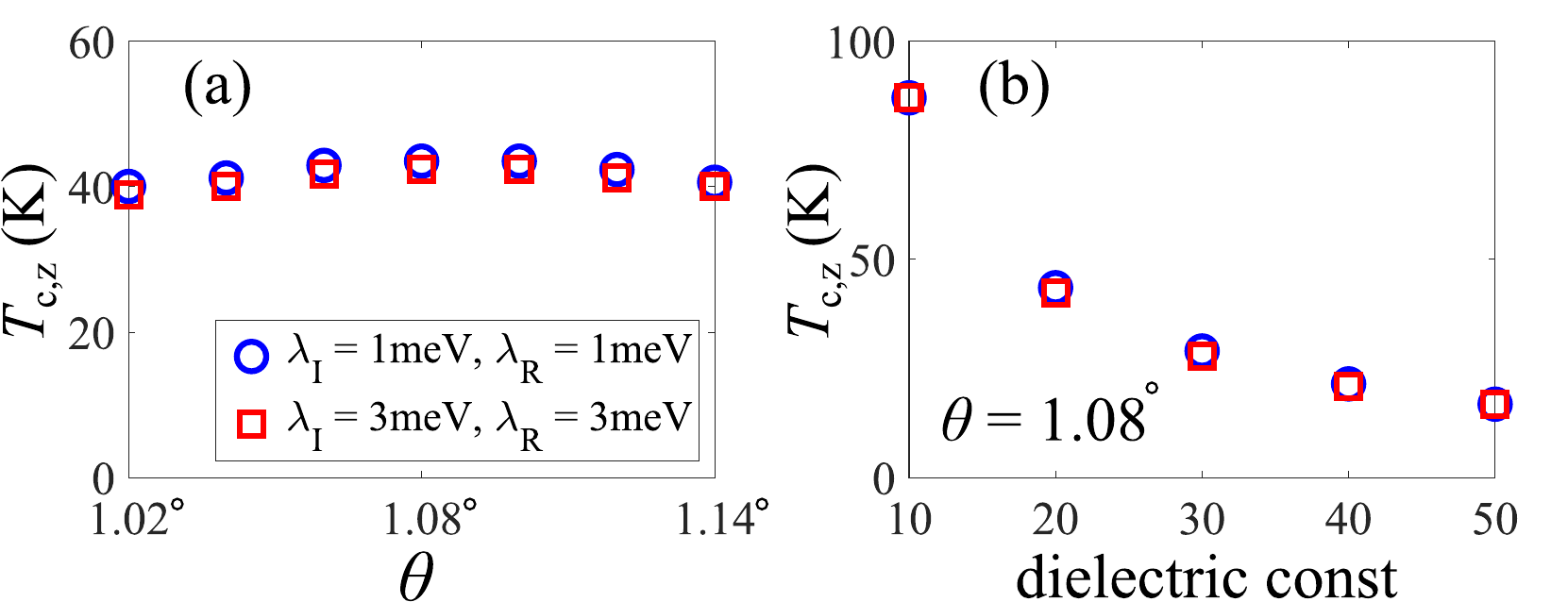}
		\caption{Transition temperature of valley polarization. We numerically solve Eqs.~(\ref{Eq:LE_m}) and (\ref{Eq:MM_z}) with a $24\times 24$ momentum grid incorporating the four moir\'e bands. (a) $T_{c,z}$ as a function of $\theta$ with $\epsilon=20$ and $d=400\text{\AA}$. (b) $T_{c,z}$ as a function of dielectric constant ($\epsilon$) with $\theta=1.08^\circ$ and $d=400\text{\AA}$.
		}
		\label{Fig:VP}
	\end{figure}

	\textit{Mean field theory and valley polarization. ---} The Coulomb interaction can significantly modify the single-particle phase diagram discussed above. We assume the system is sandwiched by two metallic gates with a half-gate distance $d$, and the screened Coulomb potential in momentum space is modeled by $\tilde{V}(\vex{p})=2\pi e^2\tanh(|\vex{p}|d)/(\epsilon|\vex{p}|)$ \cite{WuF2020,ChouYZ2022a}, where $e$ is the electron charge and $\epsilon$ is the dielectric constant.

	First, we project the microscopic model to the four moir\'e bands near charge neutrality. The projected Hamiltonian is given by $\hat{H}_{0}+\hat{H}_{I}$, where
	\begin{subequations}\label{Eq:H_proj}
		\begin{align}
			\label{Eq:H_proj_0}\hat{H}_{0}=&\sum_{\vex{k}}\sum_b\sum_{\tau}\mathcal{E}_{\tau,b}(\vex{k})\phi^{\dagger}_{\tau,b}(\vex{k})\phi_{\tau,b}(\vex{k}),\\
			\nonumber\hat{H}_{I}=&\frac{1}{2\mathcal{A}}\sum_{\vex{k}_1,\vex{k}_2,\vex{q}}\sum_{\tau_1,\tau_2}\sum_{b_1,b_2,b_3,b_4}\mathcal{V}^{(\tau_1,\tau_2),(b_1,b_2,b_3,b_4)}_{\vex{k}_1+\vex{q},\vex{k}_2-\vex{q},\vex{k}_2,\vex{k}_1}\\
			\label{Eq:H_proj_I}&\times\phi^{\dagger}_{\tau_1,b_1}(\vex{k}_1+\vex{q})\phi^{\dagger}_{\tau_2,b_2}(\vex{k}_2-\vex{q})\phi_{\tau_2,b_3}(\vex{k}_2)\phi_{\tau_1,b_4}(\vex{k}_1).
		\end{align}
	\end{subequations}
	In the above expressions, $\tau$ denotes the valley index, $\mathcal{E}_{\tau,b}(\vex{k})$ represents the energy of $b$th band in the $\tau K$valley, $\phi_{\tau,b}$ is the electron annihilation operator, $\mathcal{A}$ is the area of the 2D system, and $\mathcal{V}$ is the projected Coulomb interaction \cite{CeaT2022a,SM}.

	With the projected Hamiltonian, we develop a MF theory using the projected Coulomb interaction. In the Supplemental Material (SM) \cite{SM}, we show that the $u$-channel (in the Mandelstam variables \cite{PeskinME2018}, associated with the Fock decomposition) of the Coulomb interaction can be expressed by squares of fermion bilinears, $\hat{\Sigma}_{b,x}$, $\hat{\Sigma}_{b,y}$, and $\hat{\Sigma}_{b,z}$. The $\hat{\Sigma}_{b,x}$ and $\hat{\Sigma}_{b,y}$ terms correspond to the intervalley coherent order \cite{YouYZ2019,KangJ2019,BultinckN2020}; the $\hat{\Sigma}_{b,z}$ term corresponds to VP \cite{RepellinC2020,AlaviradY2020,WuF2020a,ChatterjeeS2020}, resulting in the occupancy imbalance between the $+K$ and $-K$ valleys. Notably, these interaction terms are negative definite with the tendency to develop instabilities related to the bilinears discussed above. We then employ the imaginary-time path integral formalism and perform Hubbard-Stratonovich decoupling of the interaction $\hat{H}_{I}'$. The order parameter (i.e., the Hubbard-Stratonovich fields), $m_{b,\alpha}$, is associated with the fermion bilinear $\hat{\Sigma}_{b,\alpha}$. In the static limit, we integrate out the fermionic fields and derive an effective action. The Landau theory can be constructed by expanding $m_{b,\alpha}$. We focus on the orders with $\vex{q}=0$ and derive the linearized equations for the transition temperature as follows (See SM \cite{SM} for detailed derivations.):
	\begin{subequations}\label{Eq:Inst}
		\begin{align}
			\label{Eq:LE_m}m_{b,\alpha}(\vex{k})=&\sum_{b',\vex{k}'}\mathcal{M}^{bb'}_{\alpha}(\vex{k},\vex{k}')\bigg|_{T=T_{c,\alpha}}m_{b',\alpha}(\vex{k}'),\\
			\nonumber\mathcal{M}^{bb'}_{\alpha=x,y}(\vex{k},\vex{k}')=&-\frac{1}{\mathcal{A}}\text{Re}\left[\mathcal{V}^{(+,+),(b'b'bb)}_{-\vex{k}',\vex{k}',\vex{k},-\vex{k}}\right]\\
			\label{Eq:MM_xy}&\times\frac{f(\mathcal{E}_{+,b'}(\vex{k}')-\mu)-f(\mathcal{E}_{+,b'}(-\vex{k}')-\mu)}{\mathcal{E}_{+,b'}(\vex{k}')-\mathcal{E}_{+,b'}(-\vex{k}')},\\
			\label{Eq:MM_z}\mathcal{M}^{bb'}_{z}(\vex{k},\vex{k}')=&-\frac{1}{\mathcal{A}}\mathcal{V}^{(+,+),(bb'bb')}_{\vex{k},\vex{k}',\vex{k},\vex{k}'}f'\left(\mathcal{E}_{+,b'}(\vex{k}')-\mu\right),
		\end{align}
	\end{subequations}
	where $\mu$ is the chemical potential, $f(\mathcal{E})=1/(1+e^{\mathcal{E}/(k_B T)})$ is the Fermi function, $k_B$ is the Boltzmann constant, $T$ is the temperature, and $f'(\mathcal{E})=\partial f(\mathcal{E})/\partial \mathcal{E}$ is the derivative of Fermi function. The critical temperature $T_{c,\alpha}$ for the $\alpha$ channel is determined by the smallest $T$ such that $\mathcal{M}^{bb'}_{\alpha}(\vex{k},\vex{k}')$ gives the largest eigenvalue of 1.
	Note that the linearized equation for the VP ($m_{z}$) is consistent with the results based on Hartree-Fock approximation \cite{WuF2020}, except that the self-energy correction to the band energies is not taken into account in this MF theory. 
	
	The linearized equation for valley polarized order depends on the density of states near $\mu$, while the intervalley coherent order measures the nesting between $\mathcal{E}_{+,b}(\vex{k})-\mu$ and $\mathcal{E}_{-,b}(\vex{k})-\mu$. With $\mathcal{V}^{(+,+),(bb'bb')}_{\vex{k},\vex{k}',\vex{k},\vex{k}'}$ and approximate $\mathcal{V}^{(+,+),(bb'bb')}_{-\vex{k}',\vex{k}',\vex{k},-\vex{k}}$ \cite{IVC}, we find that VP yields a higher $T_c$ than the intervalley coherent order \cite{SM}, consistent with the experimental findings \cite{LinJX2022,BhowmikS2023}. We focus only on the VP in this rest of the Letter \cite{HF}. 
	
	We extract $T_{c,z}$ as a function of doping for several $\theta$ and SOC parameters using Eqs.~(\ref{Eq:LE_m}) and (\ref{Eq:MM_z}). We find that $T_{c,z}$ is essentially doping-independent except for the band edges because the overall band spread is comparable to $T_{c,z}$. Thus, we focus on the $\theta$ and SOC effects. In Fig.~\ref{Fig:VP}(a), we plot the calculated $T_{c,z}$ (with $\epsilon=20$ and $d=400$\AA) as a function of $\theta$ for two representative SOC parameters at $\nu=-2$ (i.e., two holes per moir\'e unit cell). We find that $T_{c,z}$ peaks at the magic angle $\theta=1.08^\circ$ and slightly decreases for $\theta$ away from the magic angle. We also plot $T_{c,z}$ (at $\theta=1.08^\circ$) as a function of dielectric constant $\epsilon$ in Fig.~\ref{Fig:VP}(b), showing $T_{c,z}\approx16$K for $\epsilon=50$. Our results suggest that VP is the dominant instability in MATBG with induced SOCs, regardless of the model parameters \cite{HF}.
	
	%The presence of VP significantly modifies the low-temperature phase diagram. For example, the 2D time-reversal topological insulators, which are predicted based on single-particle analysis [see Fig.~\ref{Fig:Setup}(b) and Ref.~\cite{WangT2020}], are most likely absent due to the spontaneous time-reversal symmetry breaking arising from VP.
	
	\begin{figure}[t!]
		\includegraphics[width=0.45\textwidth]{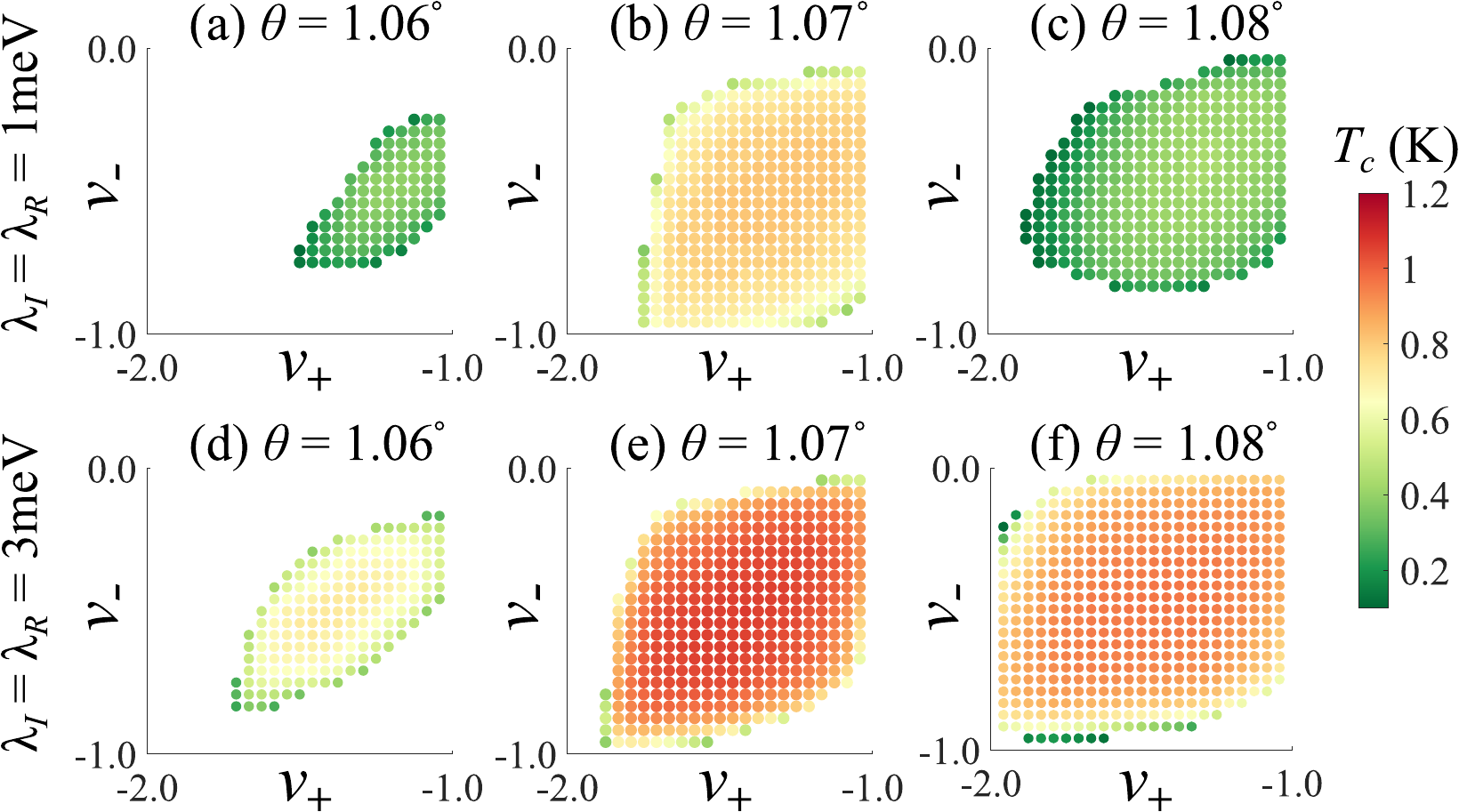}
		\caption{$T_c$ as a function of fillings in the interband ($b_+=1$ and $b_-=2$) intersublattice Ising SC. We numerically solve Eq.~(\ref{Eq:SC}) in a $24\times 24$ momentum grid with the pairing between the first band in the $+K$ valley and the second band in the $-K$ valley. We consider $\lambda_I=\lambda_R=1$meV for (a)-(c) and $\lambda_I=\lambda_R=3$meV for (d)-(f) as well as three different twist angles, $1.06^\circ$ [(a),(d)], $1.07^\circ$ [(b),(e)], and $1.08^\circ$ [(c),(f)]. The white region corresponds to $T_c<0.1$K. $\nu_+$ ($\nu_-$) is the filling factor fo the $+K$ ($-K$) valley. The Chern numbers of the active bands: $\mathcal{C}_{+,1}=1$ and $\mathcal{C}_{-,2}=1$ for (c), (f); $\mathcal{C}_{+,1}=1$ and $\mathcal{C}_{-,2}=3$ for (a), (b), (d), (e).
		}
		\label{Fig:SC}
	\end{figure}

	\textit{Superconductivity with valley imbalance. ---} Observable SC in the graphene systems is often due to intervalley pairing that arises from balanced valley populations and Fermi surface nesting. In the presence of VP, the perfect nesting is suppressed, and the absence of SC is therefore expected. Here, we challenge this usual expectation and reexamine the possibility of realizing observable SC in valley-imbalanced normal states.
	
	We consider an interband pairing scenario, specifically, pairings between the first and second (or between the third and fourth) bands. This choice of the normal state is motivated by Ref.~\cite{LinJX2022} in which only the $+K$ conduction bands are filled at $\nu=2$, suggesting a sizable effective chemical potential difference between the two valleys. In addition, we assume the acoustic-phonon mediated pairing \cite{WuF2019,ChouYZ2021,ChouYZ2022b,ChouYZ2022a} as a possible pairing mechanism. The main results here do not rely on the assumption of specific pairing mechanism.
	Within BCS approximation \cite{WuF2019,ChouYZ2021,ChouYZ2022b,ChouYZ2022a}, we derive the linearized gap equation, incorporating the valley imbalance, for superconducting transition temperature as follows \cite{SM}:
	\begin{subequations}\label{Eq:SC}
		\begin{align}
			\label{Eq:LGE_SC}\Delta^{b_+b_-}_{ss',\mathcal{X}}(\vex{k})=&\sum_{\vex{k}'}\mathcal{Q}^{b_+b_-}_{ss',\mathcal{X}}(\vex{k},\vex{k}')\bigg|_{T=T_c}\Delta^{b_+b_-}_{ss',\mathcal{X}}(\vex{k}'),\\
			\nonumber \mathcal{Q}^{b_+b_-}_{ss',\mathcal{X}}(\vex{k},\vex{k}')=&-\frac{1}{\mathcal{A}}\mathcal{G}_{ss',\mathcal{X}}^{b_+b_-b_-b_+}(\vex{k},\vex{k}')\\
			\label{Eq:Q_SC}&\times\frac{f(\mathcal{E}_{b_+}(\vex{k}')-\mu_+)-f(-\mathcal{E}_{b_-}(\vex{k}')+\mu_-)}{\mathcal{E}_{b_+}(\vex{k}')-\mu_++\mathcal{E}_{b_-}(\vex{k}')-\mu_-},
		\end{align} 
	\end{subequations}
	where $	\Delta^{b_+b_-}_{ss',\mathcal{X}}$ denotes the pairing of electrons with spin $s$ in the $b_+$th band of the $+K$ valley and spin $s'$ in the $b_-$th band of the $-K$ valley, $\mathcal{G}$ is the projected acoustic-phonon-mediated interaction, $\mathcal{X}$ indicates the intrasublattice or the intersublattice pairing, and $\mu_+$ ($\mu_-$) denotes the effective chemical potential in the $+K$ ($-K$) valley. The $T_c$ corresponds to the smallest $T$ such that Eq.~(\ref{Eq:Q_SC}) yields the largest eigenvalue of 1. For $b_+=1$, $b_-=2$ ($b_+=3$, $b_-=4$) and a positive $\lambda_I$, we find that the dominant pairing interaction $\mathcal{G}_{ss',\mathcal{X}}^{b_+b_-b_-b_+}$ corresponds to the intersublattice Ising pairing with $s,s'=\downarrow,\uparrow$ ($s,s'=\uparrow,\downarrow$), realizing a mixture of $p$-wave spin-triplet and $d$-wave spin-singlet pairings \cite{WuF2019,ChouYZ2021a,WuF2019a,ScheurerMS2020,LakeE2022,YuJ2022,WangYJ2024}. The intersublattice nature is a manifestation of the interband pairing as discussed in SM \cite{SM}. This simple analysis suggests that the \textit{mundane} acoustic-phonon mechanism can generate interband intersublattice Ising SC in partially valley polarized normal states, which is unexpected.
	
	We solve Eq.~(\ref{Eq:SC}) and extract $T_c$ treating the fillings of two valleys ($\nu_+$ for $+K$ valley and $\nu_-$ for $-K$ valley) as free parameters. In Fig.~\ref{Fig:SC}, we show that observable interband ($b_+=1$, $b_-=2$) intersublattice Ising SC with $s,s'=\downarrow,\uparrow$ can be realized for twist angles $\theta=1.06^\circ-1.08^\circ$ with two representative sets of SOC parameters. Particularly, SC with 0.1K$\le T_c\le1.2$K prevails for a wide range of $\nu_+$ and $\nu_-$ at $\theta=1.07^\circ$ and $\theta=1.08^\circ$ \cite{Coulomb}. The total Chern number of the active bands is nonzero, different from the conventional intervalley intraband pairings. We expect such SC can emerge for $\nu=\nu_++\nu_-\in(-3,-1)$. Similarly, the interband ($b_+=3$ and $b_-=4$) intersublattice SC with $s,s'=\uparrow,\downarrow$ can emerge for $\nu \in(1,3)$. The existence of observable interband SC here is due to the approximate Fermi surface nesting in the SOC-generated flat bands. A caution is that the Coulomb-induced self energy can change the band dispersion, which we leave for future studies. The sublattice structure of the wavefunctions is crucial for realizing an observable interband intersublattice superconductivity as discussed in the SM \cite{SM}. In addition, the interband SC is more likely to happen for $\theta\le 1.08^\circ$ because the spin textures do not favor interband Ising pairing for $\theta>1.08^\circ$ [e.g., Figs.~\ref{Fig:Band_angle}(c) and (f)].
	
	On the contrary, the SC arising from a valley symmetric normal state is intrasublattice Ising pairing, corresponding to a mixture of $s$-wave singlet and $f$-wave triplet. Near the magic angle, the highest $T_c$ reaches a few Kelvins \cite{SM}, which is an order of magnitude smaller than the estimated $T_{c,z}\sim 40$K of VP. 
	
	%We also study the conventional intervalley intraband SC with a valley symmetric normal state. The dominant SC is intrasublattice Ising pairing, corresponding to a mixture of $s$-wave singlet and $f$-wave triplet. Near the magic angle, we find the highest $T_c$ to be a few Kelvins, which is an order of magnitude smaller than the estimated $T_{c,z}\sim 40$K of VP. See SM \cite{SM} for a discussion.
	
	%Note that the estimated $T_c$ (for both interband and intraband SC) here does not incorporate the suppression by the Coulomb repulsion to the phonon-mediated attraction. Owing to the large density of states in the system, we expect that the Coulomb suppression is not significant (because of strong intraband screening by the carriers) and the qualitative conclusion remains unchanged.

	\textit{Discussion. ---} In the existing experiments of twisted bilayer graphene with induced SOC \cite{AroraHS2020,LinJX2022,BhowmikS2023}, only Ref.~\cite{AroraHS2020} (with a wide range of twist angles, ranging from $\theta=1.04^\circ$ to $\theta=0.79^\circ$) confirmed the existence SC, while other experiments reported VP ($\theta=0.98^\circ$ in Ref.~\cite{LinJX2022} and $\theta=1.17^\circ$ in Ref.~\cite{BhowmikS2023}). Our MF analysis suggests that VP dominates the low-temperature phases in the magic angle region. Thus, the away-from-magic-angle superconducting states in Ref.~\cite{AroraHS2020} ($\theta=0.87^\circ$ and $\theta=0.79^\circ$) are likely from valley-balanced normal states. Meanwhile, small superconducting regions next to correlated insulating states ($\nu\approx1.75$ in the $\theta=0.97^\circ$ device and $\nu\approx-2.78$ in the $\theta=1.04^\circ$ devices) were also reported in Ref.~\cite{AroraHS2020}. These correlated insulators are likely to be fully valley-polarized insulators, implying valley-imbalanced normal states for the nearby fillings. Our interband intersublattice Ising SC can provide a potential resolution for the experimentally observed SC near correlated insulators in MATBG with the presence proximity-induced SOCs. Note that the correlated insulators may not be valley polarized states in the experiment \cite{AroraHS2020}. Additional experimental examinations on the normal-state properties are needed for confirming our proposed unconventional superconducting state arising from valley imbalanced normal states.

	The interband SC in this work is conceptually similar to the idea of quantum Hall SC \cite{AkeraH1991a,NormanMR1992,MacDonaldAH1992a,NormanMR1995a,ChaudharyG2021} (which has not yet been observed). In both cases, the nearly flat bands are necessary for achieving an observable $T_c$, and Cooper pairs form in bands with a nonzero total Chern number. There are also some important differences. Particularly, the quantum Hall SC requires a sufficiently large magnetic field, while our proposed interband SC needs an interaction-induced partially valley-polarized normal state, which is likely occurring in MATBG with SOCs.

	We now mention several future directions. An interesting possibility is to investigate possible fractionalized topological phases (e.g., fractional quantum anomalous Hall effect \cite{CaiJ2023,ZengY2023,XuF2023,LuZ2023} and fractional quantum spin Hall effect \cite{KangK2024}) as the current system shares similarity with the twisted transitional metal dichalcogenide \cite{WuF2019b}.
	In addition, developing a framework analogous to the topological heavy fermion model \cite{SongZD2022a,ShiH2022a,YuJ2023b,CalugaruD2023a} is useful for understanding the emergent strongly correlated phenomena (e.g., Kondo physics \cite{ChouYZ2023,HuH2023b,HuH2023c,ZhouGD2024,ChouYZ2023e,LiY2023a,HuangC2024,DattaA2023a,LauLLH2023a,RaiG2023}) in MATBG with proximity-induced SOCs.

	\begin{acknowledgments}
		We are grateful to Andrei Bernevig, Jed Pixley, Zhentao Wang, Ming Xie, and Jihang Zhu for useful discussion. This work is supported by the Laboratory for Physical Sciences (Y.-Z.C., Y.T., and S.D.S.). F. W. is supported by National Key Research and Development Program of China (Grants No. 2022YFA1402401 and No. 2021YFA1401300) and National Natural Science Foundation of China (Grant No. 12274333).
	\end{acknowledgments}
	
	%%\bibliography{TBG_SOC}	
	
	
		\newpage \clearpage 
	
	\onecolumngrid
	
	\begin{center}
		{\large
			Topological flat bands, valley polarization, and interband superconductivity in magic-angle twisted bilayer graphene with proximitized spin-orbit couplings
			\vspace{4pt}
			\\
			SUPPLEMENTAL MATERIAL
		}
	\end{center}
	
	\setcounter{figure}{0}
	\renewcommand{\thefigure}{S\arabic{figure}}
	\setcounter{equation}{0}
	\renewcommand{\theequation}{S\arabic{equation}}
	\renewcommand{\bibnumfmt}[1]{[S#1]}
	\renewcommand{\citenumfont}[1]{S#1}

	In this supplemental material, we provide some technical details for the main results in the main text.
	
\section{Single-particle band structure}

We discuss the single-particle Hamiltonian in this section. First, we discuss the spin-orbit couplings (SOCs) in the Dirac Hamiltonian of graphene. Then, we review Bistritzer-MacDonald model and discuss our convention. 

\subsection{Dirac Hamiltonian with spin-orbit couplings}

the long-wavelength theory is given by
\begin{align}
	\hat{H}_0=v_F\int d^2\vex{x}\,\Psi^{\dagger}(\vex{x})\left[-i\tau_z\sigma_x\partial_x-i\sigma_y\partial_y\right]\Psi(\vex{x}),
\end{align}
where $\Psi^T(\vex{x})=[c_{A,+,\uparrow},c_{A,+,\downarrow},c_{B,+,\uparrow},c_{B,+,\downarrow},c_{A,-,\uparrow},c_{A,-,\downarrow},c_{B,-,\uparrow},c_{B,-,\downarrow}](\vex{x})$ is an eight-component field. We will discuss several single-particle perturbation in the following.

The non-magnetic staggered potential is described by
\begin{align}
	\hat{H}_{M}=&M\int d^2\vex{x}\Psi^{\dagger}(\vex{x})\left[\sigma_z\right]\Psi(\vex{x}).
\end{align}

The SOC terms are given by
\begin{align}
	\hat{H}_{\text{I}}=&\frac{\lambda_{\text{I}}}{2}\int d^2\vex{x}\Psi^{\dagger}(\vex{x})\left[\tau_zs_z\right]\Psi(\vex{x}),\\
	\hat{H}_{\text{R}}=&\frac{\lambda_{\text{R}}}{2}\int d^2\vex{x}\Psi^{\dagger}(\vex{x})\left[\tau_z\sigma_xs_y-\sigma_ys_x\right]\Psi(\vex{x}),\\
	\hat{H}_{\text{KM}}=&\frac{\lambda_{\text{KM}}}{2}\int d^2\vex{x}\Psi^{\dagger}(\vex{x})\left[\tau_z\sigma_zs_z\right]\Psi(\vex{x}),
\end{align}

\begin{table}[t]
	\centering
	\begin{tabular}{c||c|c|c|c|c|c}
		\toprule
		& $\hat{C}_{2z}$ & $\hat{C}_{2z,s}$ & $T_0$ & $T_s$ & $\hat{C}_{2z}T_0$ & $\hat{C}_{2z,s}T_s$ \\
		\midrule
		$\hat{H}_0$	& \checkmark & \checkmark & \checkmark  & \checkmark & \checkmark & \checkmark \\
		\midrule
		$\hat{H}_{M}$	&  $\mathsf{x}$ & $\mathsf{x}$ &  \checkmark  & \checkmark & $\mathsf{x}$ &  $\mathsf{x}$ \\
		$\hat{H}_{\text{I}}$	&  $\mathsf{x}$ & $\mathsf{x}$ &  $\mathsf{x}$  & \checkmark & \checkmark &  $\mathsf{x}$ \\
		$\hat{H}_{\text{R}}$	&  $\mathsf{x}$& \checkmark &  $\mathsf{x}$  & \checkmark & $\mathsf{x}$ &  \checkmark \\
		$\hat{H}_{\text{KM}}$	&  \checkmark & \checkmark &  $\mathsf{x}$  & \checkmark & $\mathsf{x}$ &  \checkmark \\
		
	\end{tabular}
	\caption{Summary of symmetry operation.}
	\label{tab:symmetry}
\end{table}

We consider the following symmetry operations
\begin{align}
	\hat{C}_{2z}:\,\,\, \Psi(\vex{x})\rightarrow&\tau_x\sigma_x\Psi(\mathcal{R}_{\pi}\vex{x}),\\
	\hat{C}_{2z,s}:\,\,\, \Psi(\vex{x})\rightarrow&\tau_x\sigma_xs_z\Psi(\mathcal{R}_{\pi}\vex{x}),\\
	\hat{T}_{0}:\,\,\,\Psi(\vex{x})\rightarrow&\tau_x\Psi(\vex{x}),\,\,i\rightarrow-i,\\
	\hat{T}_{s}:\,\,\,\Psi(\vex{x})\rightarrow&\tau_x(is_y)\Psi(\vex{x}),\,\,i\rightarrow-i,\\
	\hat{C}_{2z}\hat{T}_{0}:\,\,\,\Psi(\vex{x})\rightarrow&\sigma_x\Psi(\mathcal{R}_{\pi}\vex{x}),\,\,i\rightarrow-i,\\
	\hat{C}_{2z,s}\hat{T}_{s}:\,\,\,\Psi(\vex{x})\rightarrow&-\sigma_xs_x\Psi(\mathcal{R}_{\pi}\vex{x}),\,\,i\rightarrow-i.
\end{align}
The results of symmetry operations are summarized in Table~\ref{tab:symmetry}. All the SOC terms obey spinful time-reversal symmetry ($\hat{T}_s$) but not the spinless time-reversal symmetry ($\hat{T}_0$). $\hat{H}_{\text{KM}}$ obeys the $\hat{C}_{2z}$ symmetry, while $\hat{H}_{\text{I}}$ and $\hat{H}_{\text{R}}$ do not. Interestingly, $\hat{H}_{\text{I}}$ satisfies $\hat{C}_{2z}\hat{T}_{0}$ symmetry, but not $\hat{C}_{2z}$ or $\hat{T}_{0}$ symmetry.

\subsection{Bistritzer-MacDonald model}

Here, we review the construction of the Bistritzer-MacDonald model, describing the twisted bilayer graphene. The top layer of graphene is rotated by $\theta/2$, and the bottom layer of graphene is rotated by $-\theta/2$. First, the system develops a moir\'e pattern, an emergent superlattice with lattice constant $a_M=a/[2\sin(\theta/2)]$. Concomitantly, a moir\'e Brillouin zone with $K_M=|\vex{K}|[2\sin(\theta/2)]$ is formed. The top and bottom layers hybridize through the interlayer tunnelings, and the two microscopic valleys of monolayer graphene are effectively decoupled.

Following Ref.~, the Hamiltonian of $\tau=+$ can be expressed by
\begin{align}
	\hat{H}_{\text{BM}}^{(+)}=\int d\vex{x}\left[\begin{array}{cc}
		\Psi^{\dagger}_{+,t}(\vex{x}) & \Psi^{\dagger}_{+,b}(\vex{x})
	\end{array}\right]\left[\begin{array}{cc}
		\hat{U}_{\theta/2}\hat{h}^{(+)}_t(\vex{x})\hat{U}_{\theta/2}^{\dagger} & T^{\dagger}(\vex{x})\\
		T(\vex{x}) & \hat{U}_{\theta/2}^{\dagger}\hat{h}^{(+)}_b(\vex{x})\hat{U}_{\theta/2}
	\end{array}\right]\left[\begin{array}{c}
		\Psi_{+,t}(\vex{x})\\
		\Psi_{+,b}(\vex{x})
	\end{array}\right],
\end{align}
where 
\begin{align}
	\hat{h}^{(+)}_t(\vex{x})=&v_F\left[\tau_z\sigma_x(-i\partial_x-K_{t,x})+\sigma_y(-i\partial_y-K_{t,y})\right],\\
	\hat{h}^{(+)}_b(\vex{x})=&v_F\left[\tau_z\sigma_x(-i\partial_x-K_{b,x})+\sigma_y(-i\partial_y-K_{b,y})\right],
\end{align}
$\hat{U}_{\theta/2}=e^{i(\theta/4)\sigma_z}$, and $T(\vex{x})$ encodes the interlayer tunneling. This Hamiltonian can be solved by exact diagonalization in the momentum space. The kinetic energy terms are diagonal in the k space, while $T(\vex{x})$ couples electrons with different momenta, owing to the moir\'e zone folding effect. In real space, $T(\vex{x})=T_0+T_1e^{-\vex{b}_+i\cdot\vex{x}}+T_{-1}e^{-i\vex{b}_-\cdot\vex{x}}$,
where
\begin{align}
	T_0=&w_0\sigma_0+w_1\sigma_x,\\
	T_1=&w_0\sigma_0+w_1\left[\cos(2\pi/3)\sigma_x+\sin(2\pi/3)\sigma_y\right],\\
	T_{-1}=&w_0\sigma_0+w_1\left[\cos(-2\pi/3)\sigma_x+\sin(-2\pi/3)\sigma_y\right],\\
	\vex{b}_{\pm}=&\frac{4\pi}{3a_M}\left(\pm \sqrt{3}/2,3/2\right).
\end{align}
Now, we perform Fourier transform on the interlayer tunneling as follows:
\begin{align}
	\int d\vex{x}\, \Psi^{\dagger}_b(\vex{x})T(\vex{x})\Psi_{t}(\vex{x})=&\sum_{\vex{k}\in\text{1mBZ}}\sum_{\vex{G}_1,\vex{G}_2}\Psi^{\dagger}_{b,\vex{G}_1}(\vex{k})\left[T_0\delta_{\vex{G}_1,\vex{G}_2}+T_1\delta_{\vex{G}_1+\vex{b}_+,\vex{G}_2}+T_{-1}\delta_{\vex{G}_1+\vex{b}_-,\vex{G}_2}\right]\Psi_{t,\vex{G}_2}(\vex{k})\\
	\equiv&\sum_{\vex{k}\in\text{1mBZ}}\sum_{\vex{G}_1,\vex{G}_2}\Psi^{\dagger}_{b,\vex{G}_1}(\vex{k})\hat{T}_{\vex{G}_1,\vex{G}_2}\Psi_{t,\vex{G}_2}(\vex{k})
\end{align}
where $\vex{G}_1$ and $\vex{G}_2$ are the reciprocal lattice vectors of the moir\'e Brillouin zone. Specifically, $\vex{G}=\frac{4\pi}{3a_M}\left[n\left(\sqrt{3}/2,3/2\right)+m\left(-\sqrt{3}/2,3/2\right)\right]$ for integer $n$ and $m$. In our numerical calculations, we consider a $7\times 7$ grid of $\vex{G}$'s, and a larger grid does not change the band structure within our desired resolution.
We have used the Fourier transform convention:
\begin{align}
	\Psi_{t/b}(\vex{x})=\frac{1}{\sqrt{\mathcal{A}}}\sum_{\vex{k}\in\text{1mBZ}}\sum_{\vex{G}}e^{i(\vex{k}+\vex{G})\cdot\vex{x}}\Psi_{t/b,\vex{G}}(\vex{k}).
\end{align}

The Bistritzer-MacDonald of $+K$ valley is thus expressed by
\begin{align}
	\hat{H}_{\text{BM}}^{(+)}=\!\!\!\sum_{\vex{k}\in\text{1mBZ}}\sum_{\vex{G}_1,\vex{G}_2}\left[\begin{array}{cc}
		\psi^{\dagger}_{+,t,\vex{G}_1}(\vex{k}) & \psi^{\dagger}_{+,b,\vex{G}_1}(\vex{k})
	\end{array}\right]
	\hat{\mathcal{H}}^{(\tau)}(\vex{k})
	\left[\begin{array}{c}
		\psi_{+,t,\vex{G}_2}(\vex{k})\\
		\psi_{+,b,\vex{G}_2}(\vex{k})
	\end{array}\right].
\end{align}
where
\begin{align}
	\hat{\mathcal{H}}^{(\tau)}(\vex{k})=&\left[\begin{array}{cc}
		\hat{U}_{\theta/2}\hat{h}^{(+)}_t(\vex{k}+\vex{G}_1)\hat{U}_{\theta/2}^{\dagger}\delta_{\vex{G}_1,\vex{G}_2} & \hat{T}^{\dagger}_{\vex{G}_1,\vex{G}_2}\\
		\hat{T}_{\vex{G}_1,\vex{G}_2}	& \hat{U}_{\theta/2}^{\dagger}\hat{h}^{(+)}_b(\vex{k}+\vex{G}_1)\hat{U}_{\theta/2}\delta_{\vex{G}_1,\vex{G}_2}
	\end{array}\right],\\
	\psi_{+,l,\vex{G}}(\vex{k})=&\left[c_{+,\vex{G},lA\uparrow}, c_{+,\vex{G},lA\downarrow}, c_{+,\vex{G},lB\uparrow}, c_{+,\vex{G},lB\downarrow}\right]^T.
\end{align}

To incorporate the induced SOC on the top layer, we replace $\hat{h}^{(\tau)}_t(\vex{k})$ by $\hat{h}^{(\tau)}_t(\vex{k})+\frac{\lambda_R}{2}\tau s_z+\frac{\lambda_R}{2}\left(\tau\sigma_zs_y-\sigma_ys_x\right)$. The numerical procedure is the same except that the spin subspace needs to be implemented explicitly.

\section{Band projection}

The single-particle Hamiltonian can be diagonalized via $\hat{\mathcal{H}}^{(\tau)}(\vex{k})=\hat{U}^{(\tau)}(\vex{k})\hat{D}^{(\tau)}(\vex{k})\hat{U}^{(\tau),\dagger}(\vex{k})$, where 
\begin{align}
	\hat{D}^{(\tau)}(\vex{k})=\left[\begin{array}{ccc}
		\mathcal{E}_{\tau,1}(\vex{k}) & 0  & \dots\\
		0 & \mathcal{E}_{\tau,2}(\vex{k}) & \\
		\vdots & & \ddots
	\end{array}\right],\,\,\, \hat{U}^{(\tau)}=\left[\begin{array}{ccc}
		\vec{\chi}^{(\tau)}_{1}(\vex{k}) & \vec{\chi}^{(\tau)}_{1}(\vex{k}) & \dots
	\end{array}\right]
\end{align} 
are $8N_G\times8N_G$ matrices ($N_G$ is the number of $\vex{G}$ considered in the calculations) and $\vec{\chi}^{(\tau)}_{b}(\vex{k})$ is a $8N_G$ component vector. Note that the eigenvalues and eigenvectors between two valleys can be related by the time-reversal symmetry. 

The single-particle Hamiltonian is rewritten by
\begin{align}
	\hat{H}_0\rightarrow\sum_{\vex{k}\in\text{1mBZ}}\sum_{\tau}\sum_{b}\mathcal{E}_{\tau,b}(\vex{k})\phi^{\dagger}_{\tau,b}(\vex{k})\phi_{\tau,b}(\vex{k}),
\end{align}
where
\begin{align}
	\phi_{\tau,b}(\vex{k})=\sum_{\vex{G},l,\sigma,s}\chi^{(\tau)}_{b;\vex{G},l \sigma s}(\vex{k})c_{\tau,\vex{G},l\sigma s}(\vex{k}),
\end{align}

We can also express the microscopic fermions in the band basis
\begin{align}
	c_{\tau,\vex{G},l\sigma s}(\vex{k})=\sum_{b}\chi^{(\tau),*}_{b;\vex{G},l \sigma s}(\vex{k})\phi_{\tau,b}(\vex{k}).
\end{align}

\subsection{Projected Coulomb interaction}

The Coulomb interaction is given by
\begin{align}
	\hat{H}_{\text{Coulomb}}=&\frac{1}{2}\int\limits_{\vex{x}_1,\vex{x}_2}V(\vex{x}_1-\vex{x}_2):\hat{\rho}(\vex{x}_1)::\hat{\rho}(\vex{x}_2):\\
	\nonumber=&\frac{1}{2\mathcal{A}}\sum_{\vex{k}_1,\vex{k}_2,\vex{q}}'\sum_{\vex{G}_1,\vex{G}_2,\vex{G}''}\sum_{\substack{\tau_1,l_1,\sigma_1,s_1,\\ \tau_2,l_2,\sigma_2,s_2}}\tilde{V}(\vex{q}+\vex{G}'')\\
	\label{SEq:H_Coulomb}&\times:c^{\dagger}_{\tau_1,\vex{G}_1+\vex{G}'',l_1\sigma_1 s_1}(\vex{k}_1+\vex{q})\,c_{\tau_1,\vex{G}_1,l_1\sigma_1 s_1}(\vex{k}_1):
	:c^{\dagger}_{\tau_2,\vex{G}_2-\vex{G}'',l_2\sigma_2 s_2}(\vex{k}_2-\vex{q})\,c_{\tau_2,\vex{G}_2,l_2\sigma_2 s_2}(\vex{k}_2):,
\end{align}
where the sum with prime ensures that $\vex{k}_1$, $\vex{k}_2$, $\vex{k}_1+\vex{q}$, $\vex{k}_2-\vex{q}$ are in the first moir\'e Brillouin zone, $:\hat{B}:=\hat{B}-\langle \hat{B}\rangle_{\text{CNP}}$ is the normal order of $\hat{B}$ which subtracts the charge neutral configuration, we have used
\begin{align}
	V(\vex{x}_1-\vex{x}_2)=\frac{1}{\mathcal{A}}\sum_{\vex{p}}e^{i\vex{p}\cdot(\vex{x}_1-\vex{x}_2)}\tilde{V}(\vex{p}).
\end{align}

In this work, we assume that the system is sandwiched by two metallic gates, and the Coulomb potential is given by
\begin{align}
	\tilde{V}(\vex{q})=\frac{2\pi e^2\tanh(|\vex{q}|d)}{\epsilon|\vex{q}|}=\frac{2.23}{\epsilon}\frac{\tanh\left(|\vex{q}|a_0\times\frac{d}{a_0}\right)}{|\vex{q}|a_0}\text{eV.nm}^2=\frac{223}{\epsilon}\frac{\tanh\left(|\vex{q}|a_0\times\frac{d}{a_0}\right)}{|\vex{q}|a_0}\text{eV.\AA}^2,
\end{align}
where $\epsilon$ is the dielectric constant and $d$ is the vertical distance between the sample and one of the gate.

The Coulomb interaction in the mini-band basis is expressed by
\begin{align}
	\nonumber\hat{H}_I=\hat{P}\hat{H}_{\text{Coulomb}}\hat{P}=&\frac{1}{2\mathcal{A}}\sum_{\vex{k}_1,\vex{k}_2,\vex{q}}'\sum_{\vex{G}_1,\vex{G}_2,\vex{G}''}\sum_{\substack{\tau_1,l_1,\sigma_1,s_1,\\ \tau_2,l_2,\sigma_2,s_2}}\tilde{V}(\vex{q}+\vex{G}'')\\
	&\times\hat{P}:c^{\dagger}_{\tau_1,\vex{G}_1+\vex{G}'',l_1\sigma_1 s_1}(\vex{k}_1+\vex{q})\,c_{\tau_1,\vex{G}_1,l_1\sigma_1 s_1}(\vex{k}_1):
	:c^{\dagger}_{\tau_2,\vex{G}_2-\vex{G}'',l_2\sigma_2 s_2}(\vex{k}_2-\vex{q})\,c_{\tau_2,\vex{G}_2,l_2\sigma_2 s_2}(\vex{k}_2):\hat{P},\\
	=&\frac{1}{2\mathcal{A}}\sum_{\vex{k}_1,\vex{k}_2,\vex{q}}'\sum_{\tau_1,\tau_2}\sum_{b_1,b_2,b_3,b_4}\mathcal{V}^{(\tau_1,\tau_2),(b_1,b_2,b_3,b_4)}_{\vex{k}_1+\vex{q},\vex{k}_2-\vex{q},\vex{k}_2,\vex{k}_1}\phi^{\dagger}_{\tau_1,b_1}(\vex{k}_1+\vex{q})\phi^{\dagger}_{\tau_2,b_2}(\vex{k}_2-\vex{q})\phi_{\tau_2,b_3}(\vex{k}_2)\phi_{\tau_1,b_4}(\vex{k}_1),
\end{align}
where $\hat{P}$ is the many-body projector onto the mini-band Hilbert space, and $\mathcal{V}$ is the projected Coulomb interaction. For $\vex{q}\neq 0$,
\begin{align}
	\nonumber\mathcal{V}^{(\tau_1,\tau_2),(b_1,b_2,b_3,b_4)}_{\vex{k}_1+\vex{q},\vex{k}_2-\vex{q},\vex{k}_2,\vex{k}_1}=\sum_{\vex{G}''}\tilde{V}(\vex{q}+\vex{G}'')&\sum_{\vex{G}_1,\sigma_1,l_1,s_1}\left[\chi^{(\tau_1)}_{b_1;\vex{G}_1+\vex{G}'',l_1,\sigma_1,s_1}(\vex{k}_1+\vex{q})\chi^{(\tau_1),*}_{b_4;\vex{G}_1,l_1,\sigma_1,s_1}(\vex{k}_1)\right]\\
	\times&\sum_{\vex{G}_2,\sigma_2,l_2,s_2}\left[\chi^{(\tau_2)}_{b_2;\vex{G}_2-\vex{G}'',l_2,\sigma_2,s_2}(\vex{k}_2-\vex{q})\chi^{(\tau_2),*}_{b_3;\vex{G}_2,l_2,\sigma_2,s_2}(\vex{k}_2)\right].
\end{align}
For $\vex{q}=0$,
\begin{align}
	\nonumber\mathcal{V}^{(\tau_1,\tau_2),(b_1,b_2,b_3,b_4)}_{\vex{k}_1,\vex{k}_2,\vex{k}_2,\vex{k}_1}=\sum_{\vex{G}''\neq 0}\tilde{V}(\vex{G}'')&\sum_{\vex{G}_1,\sigma_1,l_1,s_1}\left[\chi^{(\tau_1)}_{b_1;\vex{G}_1+\vex{G}'',l_1,\sigma_1,s_1}(\vex{k}_1)\chi^{(\tau_1),*}_{b_4;\vex{G}_1,l_1,\sigma_1,s_1}(\vex{k}_1)\right]\\
	\times&\sum_{\vex{G}_2,\sigma_2,l_2,s_2}\left[\chi^{(\tau_2)}_{b_2;\vex{G}_2-\vex{G}'',l_2,\sigma_2,s_2}(\vex{k}_2)\chi^{(\tau_2),*}_{b_3;\vex{G}_2,l_2,\sigma_2,s_2}(\vex{k}_2)\right].
\end{align}
The $\vex{q}=0$ part of the Coulomb interaction has been subtracted by the charge neutrality background.

In this work, we need to evaluate
\begin{align}
	\nonumber\mathcal{V}^{(+,+),(b,b',b,b')}_{\vex{k},\vex{k}',\vex{k},\vex{k}'}=\sum_{\vex{G}''}\tilde{V}(\vex{k}-\vex{k}'+\vex{G}'')&\sum_{\vex{G}_1,\sigma_1,l_1,s_1}\left[\chi^{(+)}_{b;\vex{G}_1+\vex{G}'',l_1,\sigma_1,s_1}(\vex{k})\chi^{(+),*}_{b';\vex{G}_1,l_1,\sigma_1,s_1}(\vex{k}')\right]\\
	\times&\sum_{\vex{G}_2,\sigma_2,l_2,s_2}\left[\chi^{(+)}_{b';\vex{G}_2-\vex{G}'',l_2,\sigma_2,s_2}(\vex{k}')\chi^{(+),*}_{b;\vex{G}_2,l_2,\sigma_2,s_2}(\vex{k})\right],\\
	\nonumber\mathcal{V}^{(+,+),(b'b'bb)}_{-\vex{k}',\vex{k}',\vex{k},-\vex{k}}=\sum_{\vex{G}''}\tilde{V}(\vex{k}-\vex{k}'+\vex{G}'')&\sum_{\vex{G}_1,\sigma_1,l_1,s_1}\left[\chi^{(+)}_{b';\vex{G}_1+\vex{G}'',l_1,\sigma_1,s_1}(-\vex{k}')\chi^{(+),*}_{b;\vex{G}_1,l_1,\sigma_1,s_1}(-\vex{k})\right]\\
	\times&\sum_{\vex{G}_2,\sigma_2,l_2,s_2}\left[\chi^{(+)}_{b';\vex{G}_2-\vex{G}'',l_2,\sigma_2,s_2}(\vex{k}')\chi^{(+),*}_{b;\vex{G}_2,l_2,\sigma_2,s_2}(\vex{k})\right].
\end{align}
Following Ref.~\cite{CeaT2022a}, we simplify the evaluation of the matrix elements as follows: We include only the $\vex{G}$ such that $|\vex{k}_2-\vex{k}_1+\vex{G}|$ is minimized. The simplification here reduces the computational cost. Note that $\mathcal{V}^{(+,+),(b'b'bb)}_{-\vex{k}',\vex{k}',\vex{k},-\vex{k}}$ generally requires four different momenta, so the numerical diagonalization likely generates random phase factor. An approximation or gauge fixing is required for the numerical evaluation. Specifically, we compute $|\mathcal{V}^{(+,+),(b'b'bb)}_{-\vex{k}',\vex{k}',\vex{k},-\vex{k}}|$ instead of $\mathcal{V}^{(+,+),(b'b'bb)}_{-\vex{k}',\vex{k}',\vex{k},-\vex{k}}$.

\subsection{Useful relations in the projected Coulomb interaction}

The wavefunctions of two valleys are related by the spinful time-reversal symmetry as follows:
\begin{align}
	\chi^{(-),*}_{b;\vex{G},l,\sigma,s}(\vex{k})=&\sum_{s'}(i\hat{\sigma}_y)_{s,s'}\chi^{(+)}_{b;-\vex{G},l,\sigma,s'}(-\vex{k}),\\
	\sum_{\vex{G},l,\sigma,s}\chi^{(-)}_{b_1;\vex{G}+\vex{G}',l,\sigma,s}(\vex{k})\chi^{(-),*}_{b_2;\vex{G},l',\sigma',s}(\vex{k}')=&\sum_{\vex{G},l,\sigma,s}\chi^{(+),*}_{b_1;-\vex{G}-\vex{G}',l,\sigma,s}(-\vex{k})\chi^{(+)}_{b_2;-\vex{G},l',\sigma',s}(-\vex{k}')\\
	=&\sum_{\vex{G}'',l,\sigma,s}\chi^{(+)}_{b_2;\vex{G}''+\vex{G}',l',\sigma',s}(-\vex{k}')\chi^{(+),*}_{b_1;\vex{G}'',l,\sigma,s}(-\vex{k}),
\end{align}
where we have replaced $-\vex{G}-\vex{G}'$ by $\vex{G}''$ in the last line.
Note that the Pauli matrices for spins cancel with each other in the inner product expression. 
Using the general properties discussed above, we find that
\begin{align}
	&\mathcal{V}^{(\tau_1,\tau_2),(b_1,b_2,b_3,b_4)}_{\vex{k}_1+\vex{q},\vex{k}_2-\vex{q},\vex{k}_2,\vex{k}_1}=\mathcal{V}^{(-\tau_1,\tau_2),(b_4,b_2,b_3,b_1)}_{-\vex{k}_1,\vex{k}_2-\vex{q},\vex{k}_2,-\vex{k}_1-\vex{q}}
	=\mathcal{V}^{(\tau_1,-\tau_2),(b_1,b_3,b_2,b_4)}_{\vex{k}_1+\vex{q},-\vex{k}_2,-\vex{k}_2+\vex{q},\vex{k}_1}=\mathcal{V}^{(-\tau_1,-\tau_2),(b_4,b_3,b_2,b_1)}_{b_2b_4b_3b_1;-\vex{k}_1,-\vex{k}_2,-\vex{k}_2+\vex{q},-\vex{k}_1-\vex{q}},\\[2mm]
	&\left[\mathcal{V}^{(\tau_1,\tau_2),(b_1,b_2,b_3,b_4)}_{\vex{k}_1+\vex{q},\vex{k}_2-\vex{q},\vex{k}_2,\vex{k}_1}\right]^*=\mathcal{V}^{(\tau_2,\tau_1),(b_3,b_4,b_1,b_2)}_{b_4b_2b_1b_3;\vex{k}_2,\vex{k}_1,\vex{k}_1+\vex{q},\vex{k}_2-\vex{q}}.
\end{align}
Thus, all the projected Coulomb interaction matrix elements can be expressed by the $+K$ only results.

\section{Weak coupling theory for Coulomb-induced instabilities}

In this section, we develop a systematic theory of the Fermi surface instabilities using the projected Coulomb interaction.

\subsection{Interaction in the $u$-channel: Single-band case}

We consider the $b$th band projected Coulomb interaction [Eq.~(\ref{SEq:H_Coulomb})] with momentum transfer $\vex{q}\rightarrow\vex{k}_2-\vex{k}_1+\vex{q}$ (corresponding to the $u$-channel scattering in the Mandelstam variables) as follows
\begin{align}
	\hat{H}_{b,I,u\text{-channel}}=&\frac{1}{2\mathcal{A}}\sum_{\vex{k}_1,\vex{k}_2,\vex{q}}\sum_{\tau_1,\tau_2}\mathcal{V}^{(\tau_1,\tau_2)}_{\vex{k}_2+\vex{q},\vex{k}_1-\vex{q},\vex{k}_2,\vex{k}_1}\phi^{\dagger}_{\tau_1,b}(\vex{k}_2+\vex{q})\phi^{\dagger}_{\tau_2,b}(\vex{k}_1-\vex{q})\phi_{\tau_2,b}(\vex{k}_2)\phi_{\tau_1,b}(\vex{k}_1)\\
	=&\frac{-1}{2\mathcal{A}}\sum_{\vex{k}_1,\vex{k}_2,\vex{q}}\sum_{\tau_1,\tau_2}\mathcal{V}^{(\tau_1,\tau_2)}_{\vex{k}_2+\vex{q},\vex{k}_1-\vex{q},\vex{k}_2,\vex{k}_1}\phi^{\dagger}_{\tau_1,b}(\vex{k}_2+\vex{q})\phi_{\tau_2,b}(\vex{k}_2)\phi^{\dagger}_{\tau_2,b}(\vex{k}_1-\vex{q})\phi_{\tau_1,b}(\vex{k}_1)+\dots\\
	=&\frac{-1}{2\mathcal{A}}\sum_{\vex{k}_1,\vex{k}_2,\vex{q}}\left[\begin{array}{r}
		\mathcal{V}^{(+,+)}_{\vex{k}_2+\vex{q},\vex{k}_1-\vex{q},\vex{k}_2,\vex{k}_1}\phi^{\dagger}_{+,b}(\vex{k}_2)\phi_{+,b}(\vex{k}_2+\vex{q})\phi^{\dagger}_{+,b}(\vex{k}_1-\vex{q})\phi_{+,b}(\vex{k}_1)\\[1mm]
		+\mathcal{V}^{(-,-)}_{\vex{k}_2+\vex{q},\vex{k}_1-\vex{q},\vex{k}_2,\vex{k}_1}\phi^{\dagger}_{-,b}(\vex{k}_2+\vex{q})\phi_{-,b}(\vex{k}_2)\phi^{\dagger}_{-,b}(\vex{k}_1-\vex{q})\phi_{-,b}(\vex{k}_1)\\[1mm]
		+\mathcal{V}^{(+,-)}_{\vex{k}_2+\vex{q},\vex{k}_1-\vex{q},\vex{k}_2,\vex{k}_1}\phi^{\dagger}_{+,b}(\vex{k}_2+\vex{q})\phi_{-,b}(\vex{k}_2)\phi^{\dagger}_{-,b}(\vex{k}_1-\vex{q})\phi_{+,b}(\vex{k}_1)\\[1mm]
		+\mathcal{V}^{(-,+)}_{\vex{k}_2+\vex{q},\vex{k}_1-\vex{q},\vex{k}_2,\vex{k}_1}\phi^{\dagger}_{-,b}(\vex{k}_2+\vex{q})\phi_{+,b}(\vex{k}_2)\phi^{\dagger}_{+,b}(\vex{k}_1-\vex{q})\phi_{-,b}(\vex{k}_1)
	\end{array}\right]+\dots\\
	\label{SEq:Int_F_mid}=&\frac{-1}{2\mathcal{A}}\sum_{\vex{k}_1,\vex{k}_2,\vex{q}}\left[\begin{array}{r}
		\mathcal{V}^{(+,+)}_{\vex{k}_2+\vex{q},\vex{k}_1-\vex{q},\vex{k}_2,\vex{k}_1}\phi^{\dagger}_{+,b}(\vex{k}_2)\phi_{+,b}(\vex{k}_2+\vex{q})\phi^{\dagger}_{+,b}(\vex{k}_1-\vex{q})\phi_{+,b}(\vex{k}_1)\\[1mm]
		+\mathcal{V}^{(+,+)}_{\vex{k}_1,\vex{k}_2,\vex{k}_1+\vex{q},\vex{k}_2-\vex{q}}\phi^{\dagger}_{-,b}(-\vex{k}_2+\vex{q})\phi_{-,b}(-\vex{k}_2)\phi^{\dagger}_{-,b}(-\vex{k}_1-\vex{q})\phi_{-,b}(-\vex{k}_1)\\[1mm]
		+\mathcal{V}^{(+,+)}_{\vex{k}_2+\vex{q},-\vex{k}_2,\vex{k}_1+\vex{q},-\vex{k}_1}\phi^{\dagger}_{+,b}(\vex{k}_2+\vex{q})\phi_{-,b}(\vex{k}_2)\phi^{\dagger}_{-,b}(-\vex{k}_1-\vex{q})\phi_{+,b}(-\vex{k}_1)\\[1mm]
		+\mathcal{V}^{(+,+)}_{-\vex{k}_1,\vex{k}_1-\vex{q},-\vex{k}_2,\vex{k}_2-\vex{q}}\phi^{\dagger}_{-,b}(-\vex{k}_2+\vex{q})\phi_{+,b}(-\vex{k}_2)\phi^{\dagger}_{+,b}(\vex{k}_1-\vex{q})\phi_{-,b}(\vex{k}_1)
	\end{array}\right]+\dots,
\end{align}
where we have ignored the bilinear terms and suppressed the band indexes.

Before we proceed further, we discuss the interaction matrix elements. The first two terms in Eq.~(\ref{SEq:Int_F_mid}) contain $\mathcal{V}^{(+,+)}_{\vex{k}_2+\vex{q},\vex{k}_1-\vex{q},\vex{k}_2,\vex{k}_1}$ and $\mathcal{V}^{(+,+)}_{\vex{k}_1,\vex{k}_2,\vex{k}_1+\vex{q},\vex{k}_2-\vex{q}}$ interaction matrix elements, which are exactly the same in the limit $\vex{q}=0$. The last two interaction matrix elements in Eq.~(\ref{SEq:Int_F_mid}) in the limit $\vex{q}\rightarrow 0$ are related by complex conjugation. In addition, we expect that the $\mathcal{U}(\vex{q})$ part in the projected Coulomb interaction is generically nonzero. Thus, we can approximate Eq.~(\ref{SEq:Int_F_mid}) by
\begin{align}
	\hat{H}_{b,I,u\text{-channel}}\approx&\frac{-1}{2\mathcal{A}}\sum_{\vex{k}_1,\vex{k}_2,\vex{q}}\left\{\begin{array}{c}
		\mathcal{U}_{\vex{k}_1,\vex{k}_2;\vex{q}}\left[\hat{\Lambda}(\vex{k}_1;-\vex{q})\hat{\Lambda}(\vex{k}_2;\vex{q})+\hat{\Sigma}_z(\vex{k}_1;-\vex{q})\hat{\Sigma}_z(\vex{k}_2;\vex{q})\right]\\
		+\mathcal{W}_{\vex{k}_1,\vex{k}_2;\vex{q}}\left[\hat{\Sigma}_x(\vex{k}_1;-\vex{q})\hat{\Sigma}_x(\vex{k}_2;\vex{q})+\hat{\Sigma}_y(\vex{k}_1;-\vex{q})\hat{\Sigma}_y(\vex{k}_2;\vex{q})\right]\end{array}\right\},
\end{align}
where 
\begin{align}
	\hat{\Lambda}_b(\vex{k};\vex{q})=&\phi^{\dagger}_{+,b}(\vex{k}-\vex{q})\phi_{+,b}(\vex{k})+\phi^{\dagger}_{-,b}(-\vex{k}-\vex{q})\phi_{-,b}(-\vex{k}),\\
	\hat{\Sigma}_{b,x}(\vex{k};\vex{q})=&\phi^{\dagger}_{+,b}(\vex{k}-\vex{q})\phi_{-,b}(\vex{k})+\phi^{\dagger}_{-,b}(-\vex{k}-\vex{q})\phi_{+,b}(-\vex{k}),\\
	\hat{\Sigma}_{b,y}(\vex{k};\vex{q})=&-i\phi^{\dagger}_{+,b}(\vex{k}-\vex{q})\phi_{-,b}(\vex{k})+i\phi^{\dagger}_{-,b}(-\vex{k}-\vex{q})\phi_{+,b}(-\vex{k}),\\
	\hat{\Sigma}_{b,z}(\vex{k};\vex{q})=&\phi^{\dagger}_{+,b}(\vex{k}-\vex{q})\phi_{+,b}(\vex{k})-\phi^{\dagger}_{-,b}(-\vex{k}-\vex{q})\phi_{-,b}(-\vex{k}).
\end{align}
The interaction $\mathcal{U}_{\vex{k}_1,\vex{k}_2;\vex{q}=0}=\mathcal{V}^{(+,+)}_{\vex{k}_1,\vex{k}_2,\vex{k}_1,\vex{k}_2}/2\propto \tilde{V}(\vex{k}_1-\vex{k}_2)$, and $\mathcal{W}_{\vex{k}_1,\vex{k}_2;\vex{q}=0}=\text{Re}\left[\mathcal{V}^{(+,+)}_{\vex{k}_1,-\vex{k}_1,\vex{k}_2,-\vex{k}_2}\right]/2\propto \tilde{V}(\vex{k}_2+\vex{k}_1)$.
The $\hat{\Sigma}_{x}$ and $\hat{\Sigma}_{y}$ bilinears correspond to instabilities of intervalley coherent order, while the $\hat{\Sigma}_{z}$ term corresponds to valley polarization. While the $u$-channel consists of a negative term in $\hat{\Lambda}(\vex{k}_1;-\vex{q})\hat{\Lambda}(\vex{k}_2;\vex{q})$, the overall contribution (incorporating the $s$-channel) is positive, suggesting no instability. Therefore, we ignore the $\hat{\Lambda}(\vex{k}_1;-\vex{q})\hat{\Lambda}(\vex{k}_2;\vex{q})$ term. The minimal model for the interaction-induced instability of the $b$th band is given by
\begin{align}
	\hat{H}_b=&\sum_{\vex{k}\in\text{1mBZ}}\sum_{\tau}\mathcal{E}_{\tau,b}(\vex{k})\phi^{\dagger}_{\tau,b}(\vex{k})\phi_{\tau,b}(\vex{k})-\frac{1}{2\mathcal{A}}\sum_{\vex{k}_1,\vex{k}_2,\vex{q}}\left\{\begin{array}{c}
		\mathcal{W}_{\vex{k}_1,\vex{k}_2;\vex{q}}\hat{\Sigma}_x(\vex{k}_1;-\vex{q})\hat{\Sigma}_x(\vex{k}_2;\vex{q})\\[2mm]
		+\mathcal{W}_{\vex{k}_1,\vex{k}_2;\vex{q}}\hat{\Sigma}_y(\vex{k}_1;-\vex{q})\hat{\Sigma}_y(\vex{k}_2;\vex{q})\\[2mm]
		+\mathcal{U}_{\vex{k}_1,\vex{k}_2;\vex{q}}\hat{\Lambda}(\vex{k}_1;-\vex{q})\hat{\Sigma}_z(\vex{k}_1;-\vex{q})\hat{\Sigma}_z(\vex{k}_2;\vex{q})
	\end{array}\right\}
\end{align}

\subsection{Interaction in the $u$-channel: Multi-band case}

The mini bands are often separated by small direct gaps ($\sim$ order of 1meV), and the global gaps are absent in many cases. Therefore, a multi-band formalism is needed. We focus on the same types of instabilities as in the single-band case, i.e., valley polarization and intervalley coherent order.

We focus on the $u$-channel of the projected Coulomb interaction [Eq.~(\ref{SEq:H_Coulomb})] with $\vex{q}\rightarrow \vex{k}_2-\vex{k}_1+\vex{q}$ given by
\begin{align}
	\nonumber\hat{H}_I=&-\frac{1}{2\mathcal{A}}\sum_{\vex{k}_1,\vex{k}_2,\vex{q}}\sum_{\tau_1,\tau_2}\sum_{b_1,b_2,b_3,b_4}\mathcal{V}^{(\tau_1,\tau_2),(b_1,b_2,b_3,b_4)}_{\vex{k}_2+\vex{q},\vex{k}_1-\vex{q},\vex{k}_2,\vex{k}_1}\phi^{\dagger}_{\tau_1,b_1}(\vex{k}_2+\vex{q})\phi_{\tau_2,b_3}(\vex{k}_2)\phi^{\dagger}_{\tau_2,b_2}(\vex{k}_1-\vex{q})\phi_{\tau_1,b_4}(\vex{k}_1)\\
	\nonumber=&-\frac{1}{2\mathcal{A}}\sum_{\vex{k}_1,\vex{k}_2,\vex{q}}\sum_{\tau_1,\tau_2}\sum_{b_1,b_2}\mathcal{V}^{(\tau_1,\tau_2),(b_1,b_2,b_1,b_2)}_{\vex{k}_2+\vex{q},\vex{k}_1-\vex{q},\vex{k}_2,\vex{k}_1}\phi^{\dagger}_{\tau_1,b_1}(\vex{k}_2+\vex{q})\phi_{\tau_2,b_1}(\vex{k}_2)\phi^{\dagger}_{\tau_2,b_2}(\vex{k}_1-\vex{q})\phi_{\tau_1,b_2}(\vex{k}_1)+\dots\\
	\approx&-\frac{1}{2\mathcal{A}}\sum_{\vex{k}_1,\vex{k}_2,\vex{q}}\sum_{b_1,b_2}\left[\begin{array}{c}
		\mathcal{U}^{b_1,b_2}_{\vex{k}_1,\vex{k}_2;\vex{q}}\left[\hat{\Lambda}_{b_1}(\vex{k}_1;-\vex{q})\hat{\Lambda}_{b_2}(\vex{k}_2;\vex{q})+\hat{\Sigma}_{b_1,z}(\vex{k}_1;-\vex{q})\hat{\Sigma}_{b_2,z}(\vex{k}_2;\vex{q})\right]\\[2mm]
		+\mathcal{W}^{b_1,b_2}_{\vex{k}_1,\vex{k}_2;\vex{q}}\left[\hat{\Sigma}_{b_1,x}(\vex{k}_1;-\vex{q})\hat{\Sigma}_{b_2,x}(\vex{k}_2;\vex{q})+\hat{\Sigma}_{b_1,y}(\vex{k}_1;-\vex{q})\hat{\Sigma}_{b_2,y}(\vex{k}_2;\vex{q})\right]
	\end{array}\right]+\dots,
\end{align}
where $\mathcal{U}^{b_1,b_2}_{\vex{k}_1,\vex{k}_2;\vex{q}}$ and $\mathcal{W}^{b_1,b_2}_{\vex{k}_1,\vex{k}_2;\vex{q}}$ are the interaction matrix elements for the valley polarization and the intervalley coherent order, respectively. In the limit $\vex{q}\rightarrow0$,
$\mathcal{U}^{b_1,b_2}_{\vex{k}_1,\vex{k}_2;\vex{q}=0}=\mathcal{V}^{(+,+),(b_1,b_2,b_1,b_2)}_{\vex{k}_1,\vex{k}_2,\vex{k}_1,\vex{k}_2}/2$ and $\mathcal{W}^{b_1,b_2}_{\vex{k}_1,\vex{k}_2;\vex{q}=0}=\text{Re}\left[\mathcal{V}^{(+,+),(b_1,b_1,b_2,b_2)}_{\vex{k}_1,-\vex{k}_1,\vex{k}_2,-\vex{k}_2}\right]/2$. Similar to the single-band case, the $s$-channel contributes a positive $\hat{\Lambda}_{b_1}\hat{\Lambda}_{b_2}$ terms, making the overall interaction with $\hat{\Lambda}$ positive definite. Thus, the minimal model is given by $\hat{H}_{0}'+\hat{H}_{I}'$, where
\begin{subequations}
	\begin{align}
		\hat{H}_{0}=&\sum_{\vex{k}}\sum_b\sum_{\tau}\mathcal{E}_{\tau,b}(\vex{k})\phi^{\dagger}_{\tau,b}(\vex{k})\phi_{\tau,b}(\vex{k}),\\
		\hat{H}_{I}=&-\frac{1}{2\mathcal{A}}\sum_{\vex{k}_1,\vex{k}_2,\vex{q}}\sum_{b_1,b_2}\left[\begin{array}{c}
			\mathcal{U}^{b_1,b_2}_{\vex{k}_1,\vex{k}_2;\vex{q}}\left[\hat{\Sigma}_{b_1,z}(\vex{k}_1;-\vex{q})\hat{\Sigma}_{b_2,z}(\vex{k}_2;\vex{q})\right]\\[2mm]
			+\mathcal{W}^{b_1,b_2}_{\vex{k}_1,\vex{k}_2;\vex{q}}\left[\hat{\Sigma}_{b_1,x}(\vex{k}_1;-\vex{q})\hat{\Sigma}_{b_2,x}(\vex{k}_2;\vex{q})+\hat{\Sigma}_{b_1,y}(\vex{k}_1;-\vex{q})\hat{\Sigma}_{b_2,y}(\vex{k}_2;\vex{q})\right]
		\end{array}\right]
	\end{align}
\end{subequations}

\subsection{Derivation of linearized gap equation}

\begin{align}
	\nonumber\mathcal{S}=&\int d\tau\sum_{\vex{k}}\sum_{\eta=\pm}\sum_{b}\bar{\phi}_{\eta,b}(\tau,\vex{k})\left[\partial_{\tau}+\mathcal{E}_{b,\eta}(\vex{k})-\mu\right]\phi_{\eta,b}(\tau,\vex{k})\\
	&-\frac{1}{2\mathcal{A}}\int d\tau\sum_{\vex{k}_1,\vex{k}_2,\vex{q}}\sum_{b_1,b_2}\left\{\begin{array}{c}
		\mathcal{U}^{b_1,b_2}_{\vex{k}_1,\vex{k}_2;\vex{q}}\hat{\Sigma}_{b_1,z}(\tau,\vex{k}_1;-\vex{q})\hat{\Sigma}_{b_2,z}(\tau,\vex{k}_2;\vex{q})\\[2mm]
		+\mathcal{W}^{b_1,b_2}_{\vex{k}_1,\vex{k}_2;\vex{q}}\left[\hat{\Sigma}_{b_1,x}(\tau,\vex{k}_1;-\vex{q})\hat{\Sigma}_{b_2,x}(\tau,\vex{k}_2;\vex{q})+\hat{\Sigma}_{b_1,y}(\tau,\vex{k}_1;-\vex{q})\hat{\Sigma}_{b_2,y}(\tau,\vex{k}_2;\vex{q})\right]
	\end{array}
	\right\}_{\tau}\\
	\nonumber=&\int d\tau\sum_{\vex{k}}\sum_{\eta=\pm}\sum_{b}\bar{\phi}_{\eta,b}(\tau,\vex{k})\left[\partial_{\tau}+\mathcal{E}_{\eta,b}(\vex{k})-\mu\right]\phi_{\eta,b}(\tau,\vex{k})\\
	&+\frac{1}{2}\int d\tau \sum_{\vex{q}}\left\{\begin{array}{c}
		-\sum_{\alpha}\sum_{b}\sum_{\vex{k}_1}\left[m_{b,\alpha}(\tau,\vex{k}_1;-\vex{q})\hat{\Sigma}_{b,\alpha}(\tau,\vex{k}_1;\vex{q})+\hat{\Sigma}_{b,\alpha}(\tau,\vex{k}_1;-\vex{q})m_{b,\alpha}(\tau,\vex{k}_1;\vex{q})\right]\\[2mm]
		+\mathcal{A}\sum_{b_1,b_2}\sum_{\vex{k}_1,\vex{k}_2}m_{b_1,x}(\vex{k}_1;-\vex{q})\left[\mathcal{W}^{-1}\right]^{b_1,b_2}_{\vex{k}_1,\vex{k}_2;\vex{q}}m_{b_2,x}(\vex{k}_2;\vex{q})\\[2mm]
		+\mathcal{A}\sum_{b_1,b_2}\sum_{\vex{k}_1,\vex{k}_2}m_{b_1,y}(\vex{k}_1;-\vex{q})\left[\mathcal{W}^{-1}\right]^{b_1,b_2}_{\vex{k}_1,\vex{k}_2;\vex{q}}m_{b_2,y}(\vex{k}_2;\vex{q})\\[2mm]
		+\mathcal{A}\sum_{b_1,b_2}\sum_{\vex{k}_1,\vex{k}_2}m_{b_1,z}(\vex{k}_1;-\vex{q})\left[\mathcal{U}^{-1}\right]^{b_1,b_2}_{\vex{k}_1,\vex{k}_2;\vex{q}}m_{b_2,z}(\vex{k}_2;\vex{q})
	\end{array}
	\right\}_{\tau},
\end{align}
where we have used the Hubbard Stratonovich decoupling in the last line and $m_{b,\alpha}$ is the Hubbard Stratonovich field (order parameter). Note that the bands are decoupled in the quadratic fermionic theory but are coupled through the order parameters. 
With the assumption of static order parameters (i.e., $m_{b,\alpha}$ is $\tau$-independent), we integrate out the fermionic fields and derive the free energy as follows:
\begin{align}
	\nonumber\mathcal{F}\!\!=&\!-\!\frac{1}{\beta}\sum_{\omega_n,\vex{k}}\!\sum_b\!\tr\ln\!\left\{\underbrace{\left[\begin{array}{cc}
			-i\omega_n+\mathcal{E}_b(\vex{k})-\mu & 0\\
			0 & -i\omega_n+\mathcal{E}_b(-\vex{k})-\mu
		\end{array}\right]}_{\hat{G}_{b,0}^{-1}(\omega_n,\vex{k})}\!\delta_{\vex{q},0}\!-\!\underbrace{\left[\begin{array}{cc}
			m_{b,z}(\vex{k};\vex{q}) & m_{b,x}(\vex{k};\vex{q})-im_{b,y}(\vex{k};\vex{q})\\
			m_{b,x}(-\vex{k};\vex{q})+im_{b,y}(-\vex{k};\vex{q}) & -m_{b,z}(-\vex{k};\vex{q})
		\end{array}\right]}_{\hat{M}_b(\vex{k};\vex{q})}\!\right\}\\
	&+\frac{\mathcal{A}}{2}\sum_{\vex{k}_1,\vex{k}_2}\sum_{b_1,b_2}\left[\sum_{\alpha=x,y}\!\!m_{b_1,\alpha}(\vex{k}_1;-\vex{q})\left[\mathcal{W}^{-1}\right]^{b_1,b_2}_{\vex{k}_1,\vex{k}_2;\vex{q}}m_{b_2,\alpha}(\vex{k}_2;\vex{q})+\!m_{b_1,z}(\vex{k}_1;-\vex{q})\left[\mathcal{U}^{-1}\right]^{b_1,b_2}_{\vex{k}_1,\vex{k}_2;\vex{q}}m_{b_2,z}(\vex{k}_2;\vex{q})
	\right]
\end{align}
Assuming $m_a$'s are infinitesimal, we can construct a Landau theory as follows:
\begin{align}
	\nonumber\mathcal{F}\approx&\text{const}-\frac{(-1)}{\beta}\sum_{\omega_n,{\vex{k}}}\sum_b\tr\left[\hat{G}_{b,0}(\omega_n,{\vex{k}})\hat{M}_b(\vex{k};0)\right]+\frac{(-1)^2}{2\beta}\sum_{\omega_n,{\vex{k}},\vex{q}}\sum_b\tr\left[\hat{G}_{b,0}(\omega_n,\vex{k})\hat{M}_b(\vex{k};\vex{q})\hat{G}_{b,0}(\omega_n,\vex{k}+\vex{q})\hat{M}_b(\vex{k};-\vex{q})\right]\\
	&+\frac{\mathcal{A}}{2}\sum_{\vex{k}_1,\vex{k}_2}\sum_{b_1,b_2}\left[\sum_{\alpha=x,y}\!\!m_{b_1,\alpha}(\vex{k}_1;-\vex{q})\left[\mathcal{W}^{-1}\right]^{b_1,b_2}_{\vex{k}_1,\vex{k}_2;\vex{q}}m_{b_2,\alpha}(\vex{k}_2;\vex{q})+\!m_{b_1,z}(\vex{k}_1;-\vex{q})\left[\mathcal{U}^{-1}\right]^{b_1,b_2}_{\vex{k}_1,\vex{k}_2;\vex{q}}m_{b_2,z}(\vex{k}_2;\vex{q})
	\right]+\dots\\
	\nonumber=&\text{const}+\underbrace{\frac{1}{\beta}\sum_{\omega_n,\vex{k}}\sum_b\left[\frac{m_{b,z}(\vex{k})}{-i\omega_n+\mathcal{E}_b(\vex{k})-\mu}-\frac{m_{b,z}(-\vex{k})}{-i\omega_n+\mathcal{E}_b(-\vex{k})-\mu}\right]}_{=0}\\[3mm]
	\nonumber&+\frac{1}{2\beta}\sum_{\omega_n,\vex{k}}\sum_b\left\{\begin{array}{c}
		\frac{2\left[m_{b,x}(-\vex{k};-\vex{q})m_{b,x}(\vex{k};\vex{q})+m_{b,y}(-\vex{k};-\vex{q})m_{b,y}(\vex{k};\vex{q})\right]}{\left[-i\omega_n+\mathcal{E}_b(\vex{k})-\mu\right]\left[-i\omega_n+\mathcal{E}_b(-\vex{k}-\vex{q})-\mu\right]}\\[2mm]
		+\frac{m_{b,z}(\vex{k};-\vex{q})m_{b,z}(\vex{k};\vex{q})}{[-i\omega_n+\mathcal{E}_b(\vex{k})-\mu][-i\omega_n+\mathcal{E}_b(\vex{k}+\vex{q})-\mu]}+\frac{m_{b,z}(-\vex{k};-\vex{q})m_{b,z}(-\vex{k};\vex{q})}{[-i\omega_n+\mathcal{E}_b(-\vex{k})-\mu][-i\omega_n+\mathcal{E}_b(-\vex{k}-\vex{q})-\mu]}
	\end{array}
	\right\}\\[2mm]
	&+\frac{\mathcal{A}}{2}\sum_{\vex{k}_1,\vex{k}_2}\sum_{b_1,b_2}\left[\sum_{\alpha=x,y}\!\!m_{b_1,\alpha}(\vex{k}_1;-\vex{q})\left[\mathcal{W}^{-1}\right]^{b_1,b_2}_{\vex{k}_1,\vex{k}_2;\vex{q}}m_{b_2,\alpha}(\vex{k}_2;\vex{q})+\!m_{b_1,z}(\vex{k}_1;-\vex{q})\left[\mathcal{U}^{-1}\right]^{b_1,b_2}_{\vex{k}_1,\vex{k}_2;\vex{q}}m_{b_2,z}(\vex{k}_2;\vex{q})
	\right]+\dots\\[2mm]
	=&\text{const}+\frac{\mathcal{A}}{2}\sum_{\vex{k}_1,\vex{k}_2,\vex{q}}\sum_{b_1,b_2}\left\{\begin{array}{c}
		m_{b_1,x}(\vex{k}_1;-\vex{q})\left[\left[\mathcal{W}^{-1}\right]^{b_1,b_2}_{\vex{k}_1,\vex{k}_2;\vex{q}}-2\xi_{b_1}(\vex{k}_1;\vex{q};\beta)\delta_{b_1,b_2}\delta_{\vex{k}_1,-\vex{k}_2}\right]m_{b_2,x}(\vex{k}_2;\vex{q})\\[2mm]
		+m_{b_1,y}(\vex{k}_1;-\vex{q})\left[\left[\mathcal{W}^{-1}\right]^{b_1,b_2}_{\vex{k}_1,\vex{k}_2;\vex{q}}-2\xi_{b_1}(\vex{k}_1;\vex{q};\beta)\delta_{b_1,b_2}\delta_{\vex{k}_1,-\vex{k}_2}\right]m_{b_2,y}(\vex{k}_2;\vex{q})\\[2mm]
		+m_{b_1,z}(\vex{k}_1;-\vex{q})\left[\left[\mathcal{U}^{-1}\right]^{b_1,b_2}_{\vex{k}_1,\vex{k}_2;\vex{q}}-2\eta_{b_1}(\vex{k}_1;\vex{q};\beta)\delta_{b_1,b_2}\delta_{\vex{k}_1,\vex{k}_2}\right]m_{b_2,z}(\vex{k}_2;\vex{q})
	\end{array}
	\right\}+\dots,
\end{align}
where
\begin{align}
	\xi_b(\vex{k};\vex{q};\beta)=&\frac{-1}{\mathcal{A}}\frac{f(\mathcal{E}_b(\vex{k})-\mu)-f(\mathcal{E}_b(-\vex{k}-\vex{q})-\mu)}{\mathcal{E}_b(\vex{k})-\mathcal{E}_b(-\vex{k}-\vex{q})},\\
	\eta_b(\vex{k};\vex{q};\beta)=&\frac{-1}{\mathcal{A}}\frac{f(\mathcal{E}_b(\vex{k})-\mu)-f(\mathcal{E}_b(\vex{k}+\vex{q})-\mu)}{\mathcal{E}_b(\vex{k})-\mathcal{E}_b(\vex{k}+\vex{q})},
\end{align}
and $f$ is the Fermi function.

\begin{figure}[t!]
	\includegraphics[width=0.5\textwidth]{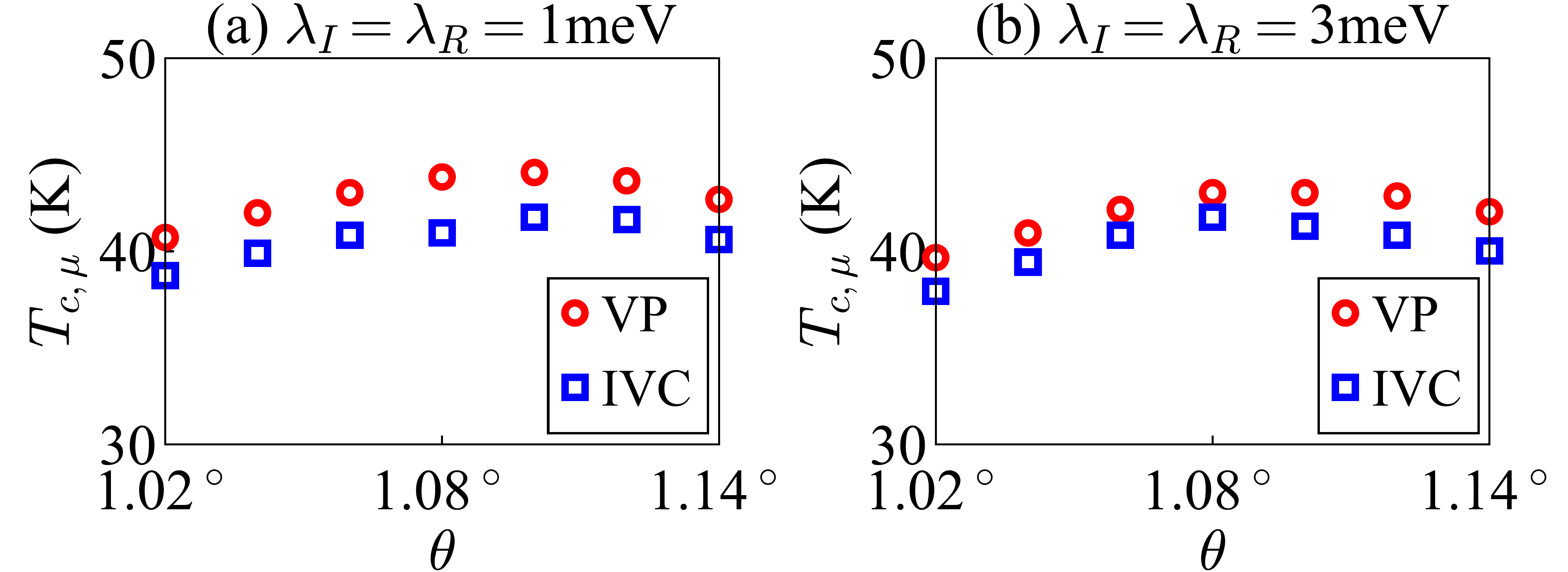}
	\caption{Transition temperatures of valley polarization and intervalley coherent order at $\nu=-2$. (a) $\lambda_I=\lambda_R=1$meV. (b) $\lambda_I=\lambda_R=3$meV. The red circles (VP) indicate the valley polarization; the blue squares (IVC) indicate the intervalley coherent order. $\epsilon=20$ and $d=400\text{\AA}$ are used.}
	\label{Fig:VP_IVC}
\end{figure}

We focus on the $\vex{q}=0$ orders. With the expression of the projected Coulomb interaction, the linearized equations are given by
\begin{align}
	m_{b,\alpha}(\vex{k};0)=&\sum_{b',\vex{k}'}\mathcal{M}^{bb'}_{\alpha}(\vex{k},\vex{k}')\bigg|_{T=T_{c,\alpha}}m_{b',\alpha}(\vex{k}';0),\\
	\mathcal{M}^{bb'}_{\alpha=x,y}(\vex{k},\vex{k}')=&-\frac{1}{\mathcal{A}}\text{Re}\left[\mathcal{V}^{(+,+),(b'b'bb)}_{-\vex{k}',\vex{k}',\vex{k},-\vex{k}}\right]\frac{f(\mathcal{E}_{+,b'}(\vex{k}')-\mu)-f(\mathcal{E}_{+,b'}(-\vex{k}')-\mu)}{\mathcal{E}_{+,b'}(\vex{k}')-\mathcal{E}_{+,b'}(-\vex{k}')},\\
	\mathcal{M}^{bb'}_{z}(\vex{k},\vex{k}')=&-\frac{1}{\mathcal{A}}\mathcal{V}^{(+,+),(bb'bb')}_{\vex{k},\vex{k}',\vex{k},\vex{k}'}f'\left(\mathcal{E}_{+,b'}(\vex{k}')-\mu\right).
\end{align}
The $T_{c,z}$ of the valley polarization can be obtained using the projected Coulomb interaction $\mathcal{V}^{(+,+),(bb'bb')}_{\vex{k},\vex{k}',\vex{k},\vex{k}'}$, which is precisely the Fock channel discussed previously.
However, the projected Coulomb interaction $\mathcal{V}^{(+,+),(bb'bb')}_{-\vex{k}',\vex{k}',\vex{k},-\vex{k}}$ is not manifestly gauge invariant because four different momenta are needed generally. Thus, approximations or \textit{ad hoc} gauge fixing is needed for the numerical evaluation. 

To examine whether valley polarization or intervalley coherent order is the dominant instability, we numerically construct $\mathcal{M}^{bb'}_{\alpha=x,y}(\vex{k},\vex{k}')$ and $\mathcal{M}^{bb'}_{z}(\vex{k},\vex{k}')$ and estimate the corresponding $T_c$. Specifically, we approximate $\mathcal{V}^{(+,+),(bb'bb')}_{-\vex{k}',\vex{k}',\vex{k},-\vex{k}}$ by $|\mathcal{V}^{(+,+),(bb'bb')}_{-\vex{k}',\vex{k}',\vex{k},-\vex{k}}|$, which overestimates the $T_{c,\alpha=x,y}$ In Fig.~\ref{Fig:VP_IVC}, we find that the $T_{c,z}$ of valley polarization is higher than the approximate $T_{c,\mu=x,y}$ of intervalley coherent order regardless of the SOC parameters and twist angle. We find very weak doping dependence in the transition temperatures, and the transition temperature hierarchy does not alter for dielectric constant $\epsilon\in[10,50]$. Thus, we conclude that valley polarization is the dominant instability within the weak coupling analysis. Nonperturbative methods like the self-consistent Hartree-Fock may predict different ground states, but this is beyond the perturbative analysis introduced in this work.

\section{Superconductivity}

We focus on in-plane acoustic-phonon-mediated pairing (ignoring frequency dependence) given by \cite{ChouYZ2022a}
\begin{align}
	\hat{H}_{\text{pairing}}=&-\frac{g}{2\mathcal{A}}\sum_{\vex{q}}\sum_l\hat{n}_l(\vex{q})\hat{n}_l(-\vex{q})\\
	=&-\frac{g}{2\mathcal{A}}\sum_{\vex{q}}\sum_l\sum_{\substack{\vex{k}_1,\vex{G}_1,\tau_1,\sigma_1,s_1\\ \vex{k}_2,\vex{G}_2,\tau_2,\sigma_2,s_2}}c^{\dagger}_{\tau_1,\vex{G}_1,l,\sigma_1,s_1}(\vex{k}_1-\vex{q})c_{\tau_1,\vex{G}_1,l,\sigma_1,s_1}(\vex{k}_1)c^{\dagger}_{\tau_2,\vex{G}_2,l,\sigma_2,s_2}(\vex{k}_2+\vex{q})c_{\tau_2,\vex{G}_2,l,\sigma_2,s_2}(\vex{k}_2),
\end{align}
where $g$ is the coupling constant. We choose $g=47.4$eV.\AA$^2$ \cite{ChouYZ2022a}. In graphene-based systems, the intervalley pairings dominate due to the nesting of time-reversal partners, and we ignore the intravalley pairings.
The intervalley pairing interaction can be approximated by a BCS Hamiltonian given by
\begin{align}
	\hat{H}_{\text{BCS}}=-\frac{g}{\mathcal{A}}\sum_{l}\sum_{\substack{\vex{k}_1,\vex{G}_1,\sigma_1,s_1\\ \vex{k}_2,\vex{G}_2,\sigma_2,s_2}}
	c^{\dagger}_{+,\vex{G}_1,l,\sigma_1,s_1}(\vex{k}_1)c^{\dagger}_{-,-\vex{G}_1,l,\sigma_2,s_2}(-\vex{k}_1)c_{-,-\vex{G}_2,l,\sigma_2,s_2}(-\vex{k}_2)c_{+,\vex{G}_2,l,\sigma_1,s_1}(\vex{k}_2),
\end{align}
where the factor of 2 in the denominator is canceled by summing of valleys. The pairing symmetry is associated with the sublattice structure as discussed in \cite{WuF2018,WuF2019,ChouYZ2021a}. Thus, the general pairing channels can be described by
\begin{align}
	\hat{H}_{\text{BCS}}=&-\frac{g}{\mathcal{A}}\sum_{l}\sum_{\substack{\vex{k}_1,\vex{G}_1,s_1\\ \vex{k}_2,\vex{G}_2,s_2}}\sum_{\sigma}
	\left[\begin{array}{r}
		c^{\dagger}_{+,\vex{G}_1,l,\sigma,s_1}(\vex{k}_1)c^{\dagger}_{-,-\vex{G}_1,l,\sigma,s_2}(-\vex{k}_1)c_{-,-\vex{G}_2,l,\sigma,s_2}(-\vex{k}_2)c_{+,\vex{G}_2,l,\sigma,s_1}(\vex{k}_2)\\
		+c^{\dagger}_{+,\vex{G}_1,l,\sigma,s_1}(\vex{k}_1)c^{\dagger}_{-,-\vex{G}_1,l,\bar\sigma,s_2}(-\vex{k}_1)c_{-,-\vex{G}_2,l,\bar\sigma,s_2}(-\vex{k}_2)c_{+,\vex{G}_2,l,\sigma,s_1}(\vex{k}_2)
	\end{array}
	\right]\\
	=&-\frac{g}{\mathcal{A}}\sum_{l}\sum_{\substack{\vex{k}_1,\vex{G}_1\\ \vex{k}_2,\vex{G}_2}}\sum_{\sigma}
	\left[\begin{array}{r}
		c^{\dagger}_{+,\vex{G}_1,l,\sigma,\uparrow}(\vex{k}_1)c^{\dagger}_{-,-\vex{G}_1,l,\sigma,\uparrow}(-\vex{k}_1)c_{-,-\vex{G}_2,l,\sigma,\uparrow}(-\vex{k}_2)c_{+,\vex{G}_2,l,\sigma,\uparrow}(\vex{k}_2)\\[2mm]
		+c^{\dagger}_{+,\vex{G}_1,l,\sigma,\uparrow}(\vex{k}_1)c^{\dagger}_{-,-\vex{G}_1,l,\sigma,\downarrow}(-\vex{k}_1)c_{-,-\vex{G}_2,l,\sigma,\downarrow}(-\vex{k}_2)c_{+,\vex{G}_2,l,\sigma,\uparrow}(\vex{k}_2)\\[2mm]
		+c^{\dagger}_{+,\vex{G}_1,l,\sigma\downarrow}(\vex{k}_1)c^{\dagger}_{-,-\vex{G}_1,l,\sigma,\uparrow}(-\vex{k}_1)c_{-,-\vex{G}_2,l,\sigma,\uparrow}(-\vex{k}_2)c_{+,\vex{G}_2,l,\sigma,\downarrow}(\vex{k}_2)\\[2mm]
		+c^{\dagger}_{+,\vex{G}_1,l,\sigma,\downarrow}(\vex{k}_1)c^{\dagger}_{-,-\vex{G}_1,l,\sigma,\downarrow}(-\vex{k}_1)c_{-,-\vex{G}_2,l,\sigma,\downarrow}(-\vex{k}_2)c_{+,\vex{G}_2,l,\sigma,\downarrow}(\vex{k}_2)\\[2mm]
		+c^{\dagger}_{+,\vex{G}_1,l,\sigma,\uparrow}(\vex{k}_1)c^{\dagger}_{-,-\vex{G}_1,l,\bar\sigma,\uparrow}(-\vex{k}_1)c_{-,-\vex{G}_2,l,\bar\sigma,\uparrow}(-\vex{k}_2)c_{+,\vex{G}_2,l,\sigma,\uparrow}(\vex{k}_2)\\[2mm]
		+c^{\dagger}_{+,\vex{G}_1,l,\sigma,\uparrow}(\vex{k}_1)c^{\dagger}_{-,-\vex{G}_1,l,\bar\sigma,\downarrow}(-\vex{k}_1)c_{-,-\vex{G}_2,l,\bar\sigma,\downarrow}(-\vex{k}_2)c_{+,\vex{G}_2,l,\sigma,\uparrow}(\vex{k}_2)\\[2mm]
		+c^{\dagger}_{+,\vex{G}_1,l,\sigma\downarrow}(\vex{k}_1)c^{\dagger}_{-,-\vex{G}_1,l,\bar\sigma,\uparrow}(-\vex{k}_1)c_{-,-\vex{G}_2,l,\bar\sigma,\uparrow}(-\vex{k}_2)c_{+,\vex{G}_2,l,\sigma,\downarrow}(\vex{k}_2)\\[2mm]
		+c^{\dagger}_{+,\vex{G}_1,l,\sigma,\downarrow}(\vex{k}_1)c^{\dagger}_{-,-\vex{G}_1,l,\bar\sigma,\downarrow}(-\vex{k}_1)c_{-,-\vex{G}_2,l,\bar\sigma,\downarrow}(-\vex{k}_2)c_{+,\vex{G}_2,l,\sigma,\downarrow}(\vex{k}_2)
	\end{array}
	\right],
\end{align}
where $\bar{A}=B$ and $\bar{B}=A$. The intrasublattice pairings correspond to $s$-wave and $f$-wave pairings, and the intersublattice pairings correspond to $p$-wave and $d$-wave pairings \cite{WuF2019,ChouYZ2021a}. In the presence of the Ising SOC, the dominant pairing is most likely of the Ising type, i.e., a mixture of spin-singlet and spin-triplet pairings. The sublattice structure depends on the normal state as will be discussed in this section.

\subsection{Single-band intraband pairings}

We consider pairing within one mini band. Thus, the BCS Hamiltonian with projected Coulomb interaction is described by
\begin{align}
	\hat{H}_{\text{BCS},b,\text{intra}}^{(s,s')}=&-\frac{1}{\mathcal{A}}\sum_{\vex{k}_1,\vex{k}_2}\mathcal{G}_{\text{intra}}^{ss'}(\vex{k}_1,\vex{k}_2)\phi^{\dagger}_{b,+}(\vex{k}_1)\phi^{\dagger}_{b,-}(-\vex{k}_1)\phi_{b,-}(-\vex{k}_2)\phi_{b,+}(\vex{k}_2),\\
	\hat{H}_{\text{BCS},b,\text{inter}}^{(s,s')}=&-\frac{1}{\mathcal{A}}\sum_{\vex{k}_1,\vex{k}_2}\mathcal{G}_{\text{inter}}^{ss'}(\vex{k}_1,\vex{k}_2)\phi^{\dagger}_{b,+}(\vex{k}_1)\phi^{\dagger}_{b,-}(-\vex{k}_1)\phi_{b,-}(-\vex{k}_2)\phi_{b,+}(\vex{k}_2),
\end{align}
where
\begin{align}
	\mathcal{G}_{\text{intra}}^{ss'}(\vex{k}_1,\vex{k}_2)=&g\sum_{l,\sigma}\sum_{\vex{G}_1,\vex{G}_2}\chi^{(+)}_{b,\vex{G}_1,l,\sigma, s}(\vex{k}_1)\chi^{(-)}_{b,-\vex{G}_1,l,\sigma,s'}(-\vex{k}_1)\chi^{(-),*}_{b,-\vex{G}_2,l,\sigma, s'}(-\vex{k}_2)\chi^{(+),*}_{b,\vex{G}_2,l,\sigma,s}(\vex{k}_2)\\
	\label{Eq:Chi_intra}=&g\sum_{l,\sigma}\sum_{\vex{G}_1,\vex{G}_2}\chi^{(+)}_{b,\vex{G}_1,l,\sigma, s}(\vex{k}_1)\chi^{(+),*}_{b,\vex{G}_1,l,\sigma,\bar{s'}}(\vex{k}_1)\chi^{(+)}_{b,\vex{G}_2,l,\sigma, \bar{s'}}(\vex{k}_2)\chi^{(+),*}_{b,\vex{G}_2,l,\sigma,s}(\vex{k}_2)\\
	\mathcal{G}_{\text{inter}}^{ss'}(\vex{k}_1,\vex{k}_2)=&g\sum_{l,\sigma}\sum_{\vex{G}_1,\vex{G}_2}\chi^{(+)}_{b,\vex{G}_1,l,\sigma, s}(\vex{k}_1)\chi^{(-)}_{b,-\vex{G}_1,l,\bar\sigma,s'}(-\vex{k}_1)\chi^{(-),*}_{b,-\vex{G}_2,l,\bar\sigma, s'}(-\vex{k}_2)\chi^{(+),*}_{b,\vex{G}_2,l,\sigma,s}(\vex{k}_2)\\
	\label{Eq:Chi_inter}=&g\sum_{l,\sigma}\sum_{\vex{G}_1,\vex{G}_2}\chi^{(+)}_{b,\vex{G}_1,l,\sigma, s}(\vex{k}_1)\chi^{(+),*}_{b,\vex{G}_1,l,\bar\sigma,\bar{s'}}(\vex{k}_1)\chi^{(+)}_{b,\vex{G}_2,l,\bar\sigma, \bar{s'}}(\vex{k}_2)\chi^{(+),*}_{b,\vex{G}_2,l,\sigma,s}(\vex{k}_2).
\end{align}
In the above expressions, we have applied the time-reversal operation in Eqs.~(\ref{Eq:Chi_intra}) and (\ref{Eq:Chi_inter}), $\bar{\uparrow}=\downarrow$, and $\bar{\downarrow}=\uparrow$.

Here, we derived the linearized gap equation using the imaginary-time path integral formalism. The imaginary-time action for the intrasublattice $ss'$ pairings is given by
\begin{align}
	\nonumber\mathcal{S}_{b,\text{intra,BCS}}^{(ss')}=&\int\limits_{\tau}\sum_{\vex{k}}\bar\phi_{+,b}(\tau,\vex{k})\left[\partial_{\tau}+\mathcal{E}_{+,b}(\vex{k})-\mu\right]\phi_{+,b}(\tau,\vex{k})+\int\limits_{\tau}\sum_{\vex{k}}\bar\phi_{-,b}(\tau,\vex{k})\left[\partial_{\tau}+\mathcal{E}_{-,b}(\vex{k})-\mu\right]\phi_{-,b}(\tau,\vex{k})\\
	&-\frac{1}{\mathcal{A}}\int\limits_{\tau}\sum_{\vex{k}_1,\vex{k}_2}\mathcal{G}_{\text{intra}}^{ss'}(\vex{k}_1,\vex{k}_2)\bar\phi_{+,b}(\tau,\vex{k}_1)\bar\phi_{-,b}(\tau,-\vex{k}_1)\phi_{-,b}(\tau,-\vex{k}_2)\phi_{+,b}(\tau,\vex{k}_2)\\
	\nonumber=&\int\limits_{\tau}\sum_{\vex{k}}\bar\phi_{+,b}(\tau,\vex{k})\left[\partial_{\tau}+\mathcal{E}_{+,b}(\vex{k})-\mu\right]\phi_{+,b}(\tau,\vex{k})+\int\limits_{\tau}\sum_{\vex{k}}\bar\phi_{-,b}(\tau,\vex{k})\left[\partial_{\tau}+\mathcal{E}_{+,b}(-\vex{k})-\mu\right]\phi_{-,b}(\tau,\vex{k})\\
	\nonumber&-\int\limits_{\tau}\sum_{\vex{k}}\left[\bar{\Delta}^{ss'}_{\text{intra}}(\tau,\vex{k})\phi_{-,b}(\tau,-\vex{k})\phi_{+,b}(\tau,\vex{k})+\bar\phi_{+,b}(\tau,\vex{k})\bar\phi_{-,b}(\tau,-\vex{k})\Delta^{ss'}_{\text{intra}}(\tau,\vex{k})\right]\\
	&+\mathcal{A}\int\limits_{\tau}\sum_{\vex{k}_1,\vex{k}_2}\left[\mathcal{G}_{\text{intra}}^{ss'}\right]^{-1}_{\vex{k}_1,\vex{k}_2}\bar{\Delta}^{ss'}_{\text{intra}}(\tau,\vex{k}_1)\Delta^{ss'}_{\text{intra}}(\tau,\vex{k}_2)
\end{align}

We are interested in the static solution, i.e., the order parameters are independent of $\tau$. The imaginary-time action becomes:
\begin{align}
	\nonumber\mathcal{S}_{b,\text{intra,BCS}}^{(ss')}\rightarrow&\frac{1}{\beta}\sum_{\omega_n}\sum_{\vex{k}}\left[\begin{array}{cc}
		\bar\phi_{+,b}(\omega_n,\vex{k}) & \phi_{-,b}(-\omega_n,-\vex{k})
	\end{array}\right]\left[\begin{array}{cc}
		-i\omega_n+\mathcal{E}_b(\vex{k})-\mu & \bar\Delta^{ss'}_{\text{intra}}(\vex{k}) \\
		\Delta^{ss'}_{\text{intra}}(\vex{k}) & -i\omega_n-\mathcal{E}_b(\vex{k})+\mu
	\end{array}\right]\left[\begin{array}{c}
		\phi_{+,b}(\omega_n,\vex{k})\\
		\bar\phi_{-,b}(-\omega_n,-\vex{k})
	\end{array}\right]\\
	&+\mathcal{A}\beta\int\limits_{\tau}\sum_{\vex{k}_1,\vex{k}_2}\left[\mathcal{G}_{\text{intra}}^{ss'}\right]^{-1}_{\vex{k}_1,\vex{k}_2}\bar{\Delta}^{ss'}_{\text{intra}}(\vex{k}_1)\Delta^{ss'}_{\text{intra}}(\vex{k}_2).
\end{align}

Then, we integrate out the fermionic fields and derive the free energy as follows:
\begin{align}
	\nonumber\mathcal{F}=&-\frac{1}{\beta}\sum_{\omega_n,\vex{k}}\ln\left\{\left[-i\omega_n+\mathcal{E}_{b}(\vex{k})-\mu\right]\left[-i\omega_n-\mathcal{E}_{b}(\vex{k})+\mu\right]+\left|\Delta^{ss'}_{\text{intra}}(\vex{k})\right|^2
	\right\}\\
	&+\mathcal{A}\sum_{\vex{k}_1,\vex{k}_2}\left[\left(\mathcal{G}_{\text{intra}}^{ss'}\right)^{-1}\right]_{\vex{k}_1,\vex{k}_2}\bar{\Delta}^{ss'}_{\text{intra}}(\vex{k}_1)\Delta^{ss'}_{\text{intra}}(\vex{k}_2).
\end{align}
Near $T_c$, the order parameters are infinitesimal, and we can derive Landau free energy as follows:
\begin{align}
	\mathcal{F}\approx&\frac{1}{\beta}\sum_{\omega_n,\vex{k}}\frac{\left|\Delta^{ss'}_{\text{intra}}(\vex{k})\right|^2}{\left[-i\omega_n+\mathcal{E}_{b}(\vex{k})-\mu\right]\left[-i\omega_n-\mathcal{E}_{b}(\vex{k})+\mu\right]}+\mathcal{A}\sum_{\vex{k}_1,\vex{k}_2}\left[\left(\mathcal{G}_{\text{intra}}^{ss'}\right)^{-1}\right]_{\vex{k}_1,\vex{k}_2}\bar{\Delta}^{ss'}_{\text{intra}}(\vex{k}_1)\Delta^{ss'}_{\text{intra}}(\vex{k}_2)\\
	=&-\sum_{\vex{k}}\frac{\tanh\left(\frac{\mathcal{E}_b(\vex{k})-\mu}{2T}\right)}{2\left[\mathcal{E}_b(\vex{k})-\mu\right]}\left|\Delta^{ss'}_{\text{intra}}(\vex{k})\right|^2
	+\mathcal{A}\sum_{\vex{k}_1,\vex{k}_2}\left[\left(\mathcal{G}_{\text{intra}}^{ss'}\right)^{-1}\right]_{\vex{k}_1,\vex{k}_2}\bar{\Delta}^{ss'}_{\text{intra}}(\vex{k}_1)\Delta^{ss'}_{\text{intra}}(\vex{k}_2).
\end{align}
The $T_c$ is determined by the vanishing quadrature, corresponding to
\begin{align}
	\Delta^{ss'}_{\text{intra}}(\vex{k})=\sum_{\vex{k}'}\left[\frac{\mathcal{G}_{\text{intra}}^{ss'}(\vex{k},\vex{k}')}{\mathcal{A}}\frac{\tanh\left(\frac{\mathcal{E}_b(\vex{k}')-\mu}{2T}\right)}{2\left[\mathcal{E}_b(\vex{k}')-\mu\right]}\right]\Delta^{ss'}_{\text{intra}}(\vex{k}').
\end{align}

\subsection{Multi-band intraband pairings}

The general multiband BCS Hamiltonian is given by
\begin{align}
	\hat{H}_{\text{BCS},b,\text{intra}}^{(s,s'),b_1,b_2,b_3,b_4}=&-\frac{1}{\mathcal{A}}\sum_{\vex{k}_1,\vex{k}_2}\mathcal{G}_{\text{intra}}^{ss',b_1b_2b_3b_4}(\vex{k}_1,\vex{k}_2)\phi^{\dagger}_{+,b_1}(\vex{k}_1)\phi^{\dagger}_{-,b_2}(-\vex{k}_1)\phi_{-,b_3}(-\vex{k}_2)\phi_{+,b_4}(\vex{k}_2),\\
	\hat{H}_{\text{BCS},b,\text{inter}}^{(s,s'),b_1,b_2,b_3,b_4}=&-\frac{1}{\mathcal{A}}\sum_{\vex{k}_1,\vex{k}_2}\mathcal{G}_{\text{inter}}^{ss',b_1b_2b_3b_4}(\vex{k}_1,\vex{k}_2)\phi^{\dagger}_{+,b_1}(\vex{k}_1)\phi^{\dagger}_{-,b_2}(-\vex{k}_1)\phi_{-,b_3}(-\vex{k}_2)\phi_{+,b_4}(\vex{k}_2).
\end{align}
For intraband pairings, we consider $b_1=b_2$ and $b_3=b_4$. We assuming valley symmetric normal states, i.e., the absence of valley polarization. The $T_c$ is determined by the linearized equation as follows:
\begin{align}
	\Delta^{ss'}_{\text{intra},b}(\vex{k})=\sum_{\vex{k}'}\sum_{b'}\left[\frac{\mathcal{G}_{\text{intra}}^{ss',bbb'b'}(\vex{k},\vex{k}')}{\mathcal{A}}\frac{\tanh\left(\frac{\mathcal{E}_{b'}(\vex{k}')-\mu}{2T}\right)}{2\left[\mathcal{E}_{b'}(\vex{k}')-\mu\right]}\right]\Delta^{ss'}_{\text{intra},b'}(\vex{k}').
\end{align}
where
\begin{align}
	\mathcal{G}_{\text{intra}}^{ss',b_1b_1b_2b_2}(\vex{k}_1,\vex{k}_2)=&g\sum_{l,\sigma}\sum_{\vex{G}_1,\vex{G}_2}\chi^{(+)}_{b_1,\vex{G}_1,l,\sigma, s}(\vex{k}_1)\chi^{(+),*}_{b_1,\vex{G}_1,l,\sigma,\bar{s'}}(\vex{k}_1)\chi^{(+)}_{b_2,\vex{G}_2,l,\sigma, \bar{s'}}(\vex{k}_2)\chi^{(+),*}_{b_2,\vex{G}_2,l,\sigma,s}(\vex{k}_2)\\
	\mathcal{G}_{\text{inter}}^{ss',b_1b_1b_2b_2}(\vex{k}_1,\vex{k}_2)=&g\sum_{l,\sigma}\sum_{\vex{G}_1,\vex{G}_2}\chi^{(+)}_{b_1,\vex{G}_1,l,\sigma, s}(\vex{k}_1)\chi^{(+),*}_{b_1,\vex{G}_1,l,\bar\sigma,\bar{s'}}(\vex{k}_1)\chi^{(+)}_{b_2,\vex{G}_2,l,\bar\sigma, \bar{s'}}(\vex{k}_2)\chi^{(+),*}_{b_2,\vex{G}_2,l,\sigma,s}(\vex{k}_2).
\end{align}

\begin{figure}[t!]
	\includegraphics[width=0.35\textwidth]{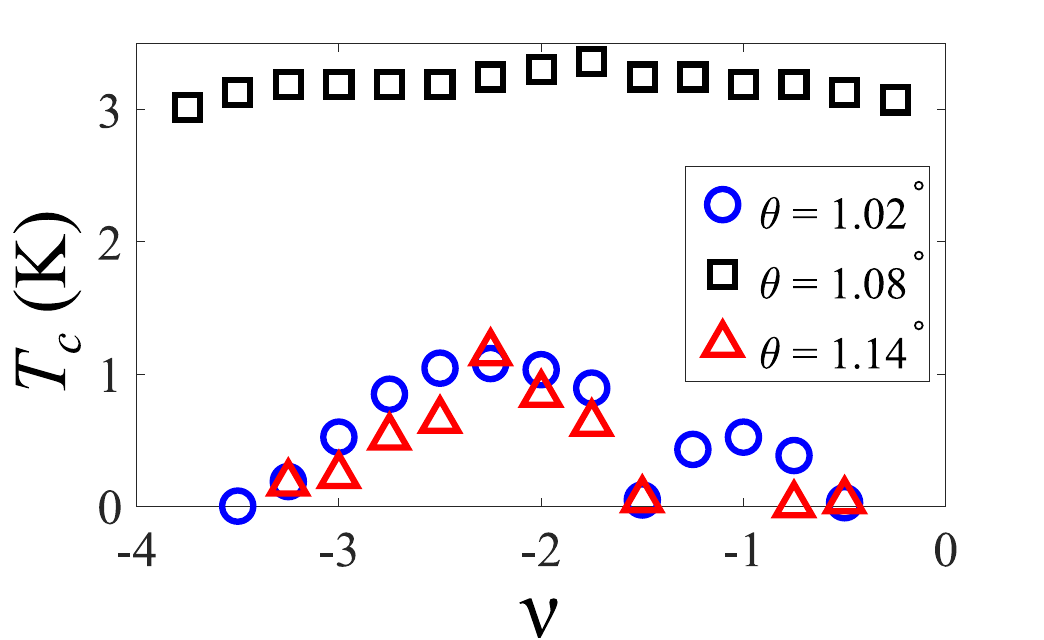}
	\caption{$T_c$ of intraband intrasublattice Ising superconductivity. The results with three different twist angles are presented. We use $\lambda_I=\lambda_R=3$meV for all the plots. }
	\label{Fig:SC_intraband}
\end{figure}

Similar to the single-band case, we find the intrasublattice Ising pairing is the leading superconductivity instability. In Fig.~\ref{Fig:SC_intraband}, we plot the $T_c$ of the intraband intrasublattice Ising superconductivity with $(s,s')=(\downarrow,\uparrow)$ for $\nu<0$. Similar results for $(s,s')=(\uparrow,\downarrow)$ can be found for $\nu>0$. For $\theta=1.08^\circ$, superconductivity prevail for a wide range of doping, and the highest $T_c$ is above 3K. For $\theta=1.02^\circ$ and $\theta=1.14^\circ$, the highest $T_c$ drops to around 1K, and the superconducting region also shrinks.

\subsection{Interband pairings}

In additional to intraband pairings, the interband pairings is also possible due to significant valley imbalance, resulting in different chemical potentials in two valleys. For simplicity, we consider the intrasublattice pairing between the $b_+$ and $b_-$ bands as follows:
\begin{align}
	\nonumber\mathcal{S}_{\mathcal{X},\text{BCS}}^{(ss'),b_+b_-b_-b_+}=&\int\limits_{\tau}\sum_{\vex{k}}\bar\phi_{+,b_+}(\tau,\vex{k})\left[\partial_{\tau}+\mathcal{E}_{+,b_+}(\vex{k})-\mu_+\right]\phi_{+,b_+}(\tau,\vex{k})+\int\limits_{\tau}\sum_{\vex{k}}\bar\phi_{-,b_-}(\tau,\vex{k})\left[\partial_{\tau}+\mathcal{E}_{-,b_-}(\vex{k})-\mu_-\right]\phi_{-,b_-}(\tau,\vex{k})\\
	&-\frac{1}{\mathcal{A}}\int\limits_{\tau}\sum_{\vex{k}_1,\vex{k}_2}\mathcal{G}_{\mathcal{X}}^{ss',b_+b_-b_-b_+}(\vex{k}_1,\vex{k}_2)\phi^{\dagger}_{+,b_+}(\tau,\vex{k}_1)\phi^{\dagger}_{-,b_-}(\tau,-\vex{k}_1)\phi_{-,b_-}(\tau,-\vex{k}_2)\phi_{+,b_+}(\tau,\vex{k}_2)\\
	\nonumber=&\int\limits_{\tau}\sum_{\vex{k}}\bar\phi_{+,b_+}(\tau,\vex{k})\left[\partial_{\tau}+\mathcal{E}_{+,b_+}(\vex{k})-\mu_+\right]\phi_{+,b_+}(\tau,\vex{k})+\int\limits_{\tau}\sum_{\vex{k}}\bar\phi_{-,b_-}(\tau,\vex{k})\left[\partial_{\tau}+\mathcal{E}_{+,b_-}(-\vex{k})-\mu_-\right]\phi_{-,b_-}(\tau,\vex{k})\\
	\nonumber&-\int\limits_{\tau}\sum_{\vex{k}}\left[\bar{\Delta}^{ss',b_+b_-}_{\mathcal{X}}(\tau,\vex{k})\phi_{-,b_-}(\tau,-\vex{k})\phi_{+,b_+}(\tau,\vex{k})+\phi_{+,b_+}^{\dagger}(\tau,\vex{k})\phi_{-,b_-}^{\dagger}(\tau,-\vex{k})\Delta^{ss',b_+b_-}_{\mathcal{X}}(\tau,\vex{k})\right]\\
	&+\mathcal{A}\int\limits_{\tau}\sum_{\vex{k}_1,\vex{k}_2}\left[\left(\mathcal{G}_{\mathcal{X}}^{ss',b_+b_-b_-b_+}\right)^{-1}\right]_{\vex{k}_1,\vex{k}_2}\bar{\Delta}^{ss',b_+b_-}_{\mathcal{X}}(\tau,\vex{k}_1)\Delta^{ss',b_+b_-}_{\mathcal{X}}(\tau,\vex{k}_2).
\end{align}
where $\mathcal{X}=$intra or inter denotes the intrasublattice or intersublattice pairing, and
\begin{align}
	\mathcal{G}_{\text{intra}}^{ss',b_+b_-b_-b_+}(\vex{k}_1,\vex{k}_2)=&g\sum_{l,\sigma}\sum_{\vex{G}_1,\vex{G}_2}\chi^{(+)}_{b_+,\vex{G}_1,l,\sigma, s}(\vex{k}_1)\chi^{(+),*}_{b_-,\vex{G}_1,l,\sigma,\bar{s'}}(\vex{k}_1)\chi^{(+)}_{b_-,\vex{G}_2,l,\sigma, \bar{s'}}(\vex{k}_2)\chi^{(+),*}_{b_+,\vex{G}_2,l,\sigma,s}(\vex{k}_2)\\
	\mathcal{G}_{\text{inter}}^{ss',b_+b_-b_-b_+}(\vex{k}_1,\vex{k}_2)=&g\sum_{l,\sigma}\sum_{\vex{G}_1,\vex{G}_2}\chi^{(+)}_{b_+,\vex{G}_1,l,\sigma, s}(\vex{k}_1)\chi^{(+),*}_{b_-,\vex{G}_1,l,\bar\sigma,\bar{s'}}(\vex{k}_1)\chi^{(+)}_{b_-,\vex{G}_2,l,\bar\sigma, \bar{s'}}(\vex{k}_2)\chi^{(+),*}_{b_+,\vex{G}_2,l,\sigma,s}(\vex{k}_2).
\end{align}

\begin{align}
	\nonumber\mathcal{S}_{\mathcal{X},\text{BCS}}^{(ss'),b_+b_-b_-b_+}\!\!\!\rightarrow&\frac{1}{\beta}\!\sum_{\omega_n}\sum_{\vex{k}}\left[\begin{array}{cc}
		\bar\phi_{+,b_+}(\omega_n,\vex{k}) & \phi_{-,b_-}(-\omega_n,-\vex{k})
	\end{array}\right]\!\!\left[\begin{array}{cc}
		-i\omega_n+\mathcal{E}_{+,b_+}(\vex{k})-\mu_+ & \bar\Delta^{ss',b_+b_-}_{\mathcal{X}}(\vex{k}) \\
		\Delta^{ss',b_+b_-}_{\mathcal{X}}(\vex{k}) & -i\omega_n-\mathcal{E}_{+,b_-}(\vex{k})+\mu_-
	\end{array}\right]\!\!\left[\begin{array}{c}
		\phi_{+,b_+}(\omega_n,\vex{k})\\
		\bar\phi_{-,b_-}(-\omega_n,-\vex{k})
	\end{array}\right]\\
	&+\mathcal{A}\beta\int\limits_{\tau}\sum_{\vex{k}_1,\vex{k}_2}\left[\left(\mathcal{G}_{\mathcal{X}}^{ss',b_+b_-b_-b_+}\right)^{-1}\right]_{\vex{k}_1,\vex{k}_2}\bar\Delta^{ss',b_+b_-}_{\mathcal{X}}(\vex{k}_1)\Delta^{ss',b_+b_-}_{\mathcal{X}}(\vex{k}_2).
\end{align}
Then, we integrate out the fermionic fields and derive the free energy as follows:
\begin{align}
	\nonumber\mathcal{F}=&-\frac{1}{\beta}\sum_{\omega_n,\vex{k}}\ln\left\{\left[-i\omega_n+\mathcal{E}_{b_+}(\vex{k})-\mu_+\right]\left[-i\omega_n-\mathcal{E}_{b_-}(\vex{k})+\mu_-\right]+\left|\Delta^{ss',b_+b_-}_{\mathcal{X}}(\vex{k})\right|^2
	\right\}\\
	&+\mathcal{A}\sum_{\vex{k}_1,\vex{k}_2}\left[\left(\mathcal{G}_{\mathcal{X}}^{ss',b_+b_-b_-b_+}\right)^{-1}\right]_{\vex{k}_1,\vex{k}_2}\bar\Delta^{ss',b_+b_-}_{\mathcal{X}}(\vex{k}_1)\Delta^{ss',b_+b_-}_{\mathcal{X}}(\vex{k}_2).
\end{align}
Near $T_c$, the order parameters are infinitesimal, and we can derive Landau free energy as follows:
\begin{align}
	\mathcal{F}\approx&\frac{1}{\beta}\sum_{\omega_n,\vex{k}}\frac{\left|\Delta^{ss',b_+b_-}_{\mathcal{X}}(\vex{k})\right|^2}{\left[-i\omega_n+\mathcal{E}_{b_+}(\vex{k})-\mu_+\right]\left[-i\omega_n-\mathcal{E}_{b_-}(\vex{k})+\mu_-\right]}+\mathcal{A}\sum_{\vex{k}_1,\vex{k}_2}\left[\left(\mathcal{G}_{\mathcal{X}}^{ss',b_+b_-b_-b_+}\right)^{-1}\right]_{\vex{k}_1,\vex{k}_2}\bar{\Delta}^{ss'}_{\mathcal{X}}(\vex{k}_1)\Delta^{ss'}_{\mathcal{X}}(\vex{k}_2)\\
	=&\sum_{\vex{k}}\frac{f(\mathcal{E}_{b_+}(\vex{k})-\mu_+)-f(-\mathcal{E}_{b_-}(\vex{k})+\mu_-)}{\mathcal{E}_{b_+}(\vex{k})-\mu_++\mathcal{E}_{b_-}(\vex{k})-\mu_-}\left|\Delta^{ss',b_+b_-}_{\mathcal{X}}(\vex{k})\right|^2
	+\mathcal{A}\sum_{\vex{k}_1,\vex{k}_2}\left[\left(\mathcal{G}_{\mathcal{X}}^{ss',b_+b_-b_-b_+}\right)^{-1}\right]_{\vex{k}_1,\vex{k}_2}\bar\Delta^{ss',b_+b_-}_{\mathcal{X}}(\vex{k}_1)\Delta^{ss',b_+b_-}_{\mathcal{X}}(\vex{k}_2).
\end{align}
The $T_c$ is determined by the vanishing quadrature, corresponding to
\begin{align}
	\Delta^{ss',b_+b_-}_{\mathcal{X}}(\vex{k})=\sum_{\vex{k}'}\left[\frac{\mathcal{G}_{\mathcal{X}}^{ss',b_+b_-b_-b_+}(\vex{k},\vex{k}')}{\mathcal{A}}\left(-\frac{f(\mathcal{E}_{b_+}(\vex{k}')-\mu_+)-f(-\mathcal{E}_{b_-}(\vex{k}')+\mu_-)}{\mathcal{E}_{b_+}(\vex{k}')-\mu_++\mathcal{E}_{b_-}(\vex{k}')-\mu_-}\right)_{T=T_c}\right]\Delta^{ss',b_+b_-}_{\mathcal{X}}(\vex{k}').
\end{align}
Note that the conventional $\tanh(\mathcal{E}/2)/(2\mathcal{E})$ kernel is recovered when setting $b_+=b_-$ and $\mu_+=\mu_-$. In our case, the above formula measures the nesting of bands $b_+$ and $b_-$ within the energy resolution $\sim T_c$. Thus, we expect such interband superconductivity emerges near the magic angle where bands are sufficiently flat. In addition, the wavefunction composition also plays an import role, which we discuss in the next subsection.

We briefly comment on the role of SOC for the interband superconductivity proposed in this work. The main consequence of having SOCs is to create strong valley imbalance that the effective chemical potentials sit at two different bands. Such a valley imbalanced normal state is required for the interband superconductivity, and MATBG with SOC can naturally realize such a normal state. We expect interband superconductivity happens in other systems with strong valley imbalanced normal states.

\subsection{Layer-sublattice-spin structure and interband superconductivity}

\begin{figure}[t!]
	\includegraphics[width=0.8\textwidth]{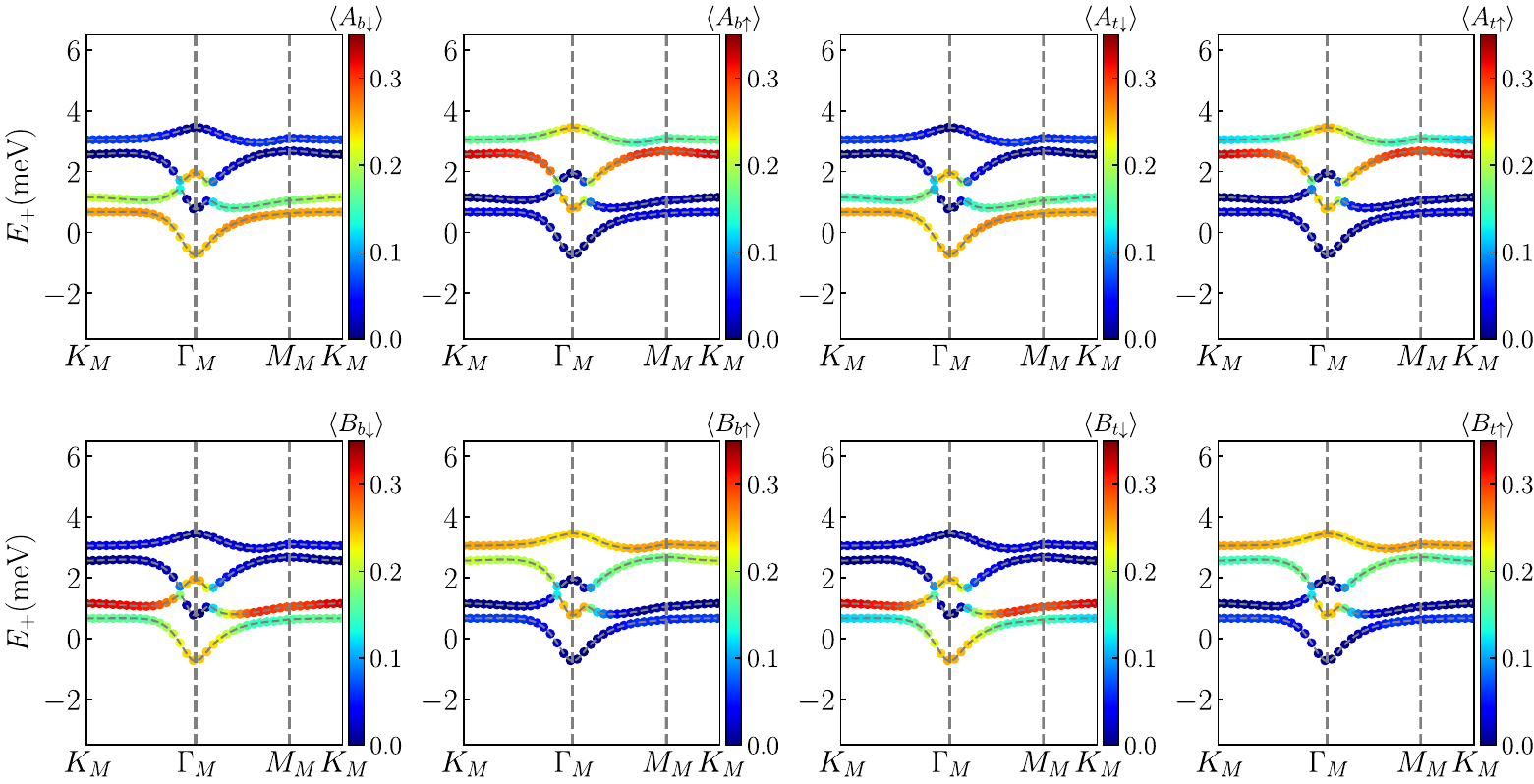}
	\caption{Layer-spin-sublattice distribution for $\theta=1.07^{\circ}$ and $\lambda_I=\lambda_R=3$meV. The mini bands in the $+K$ valley are plotted. Each subfigure shows the weight of a given layer, sublattice, and spin. $A_{ls}$ ($B_{ls}$) denotes $A$ ($B$) sublattice with layer $l$ and spin $s$.}
	\label{Fig:LS_weights_107}
\end{figure}

Here, we discuss the layer-sublattice-spin composition of the wavefunction of MATBG with SOCs. For definiteness, we focus on mini bands in the $+K$ valley. First, we investigate the $\theta=1.07^\circ$ case with $\lambda_I=\lambda_R=3$meV, which yields the highest $T_c$ for interband superconductivity. In Fig.~\ref{Fig:LS_weights_107}, we plot each layer-sublattice-spin component for the mini bands. $A_{ls}$ ($B_{ls}$) denotes $A$ ($B$) sublattice with layer $l$ and spin $s$. First, the lower (upper) two bands are mostly spin down (up), consistent with the Ising SOC. The lowest mini band is mainly governed by the sublattice $A$ ($\sim 60$ percent near $K_M$), and the second lowest mini band is mainly governed by the sublattice $B$ ($\sim 64$ percent near $K_M$). These are consistent with the intersublattice nature of the interband superconductivity. Note that the observable superconductivity happens for the effective chemical potentials near the ``flat'' part of the dispersion (i.e., the flat part around $K_M$ point). There are small differences between two layers which cannot be seen in our false color scheme. The results here support the observable interband intersublattice Ising superconductivity.

\begin{figure}[t!]
	\includegraphics[width=0.8\textwidth]{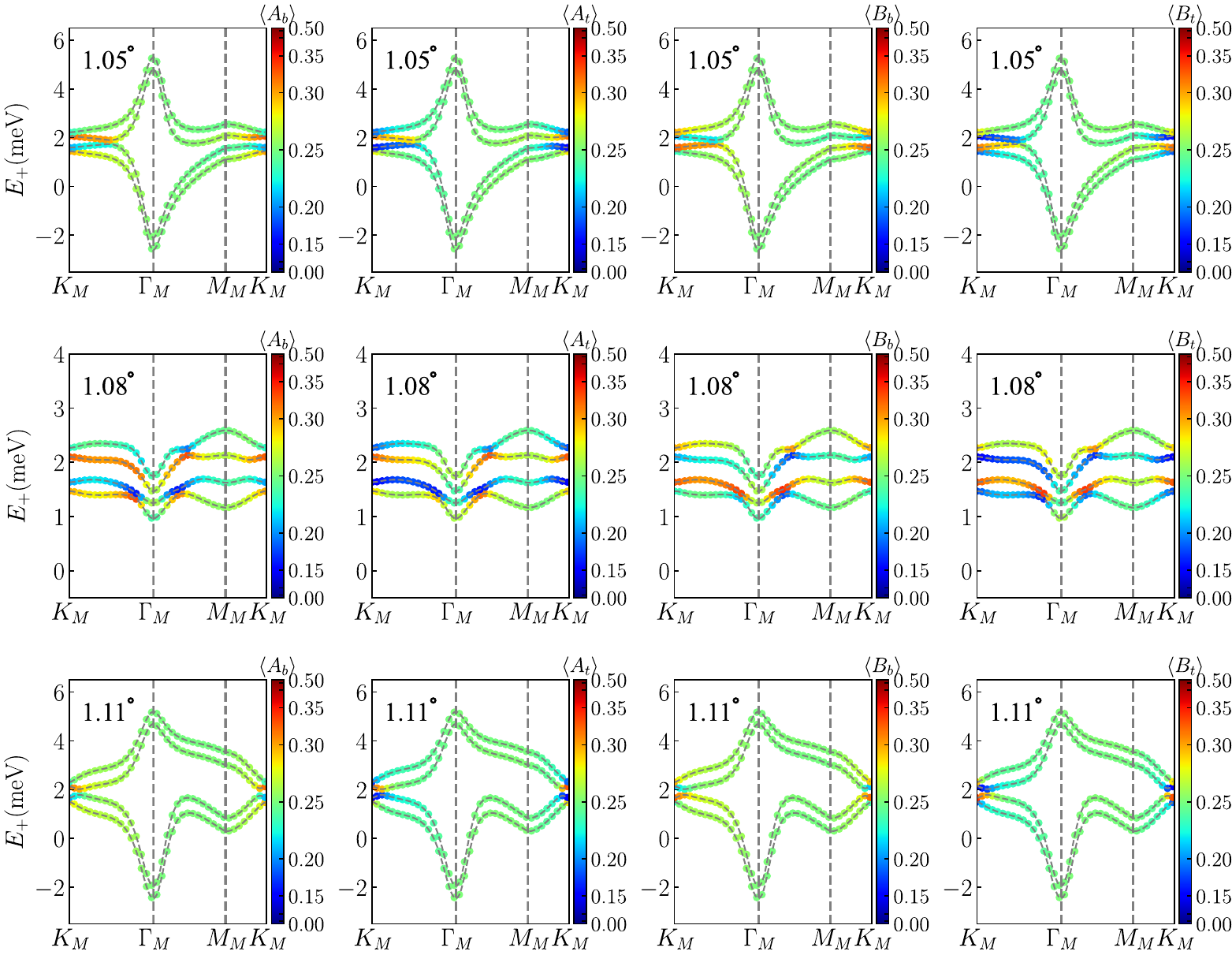}
	\caption{Layer-sublattice distribution for $\lambda_I=\lambda_R=1$meV. The mini bands in the $+K$ valley are plotted. $A_b$, $A_t$, $B_b$, and $B_t$ denote bottom layer sublattice $A$, top layer sublattice $A$, bottom layer sublattice $B$, and top layer sublattice $B$, respectively. Top row: $\theta=1.05^{\circ}$. Middle row: $\theta=1.08^{\circ}$. Bottom row: $\theta=1.11^{\circ}$}
	\label{Fig:LS_weights_I1_R1}
\end{figure}

\begin{figure}[t!]
	\includegraphics[width=0.8\textwidth]{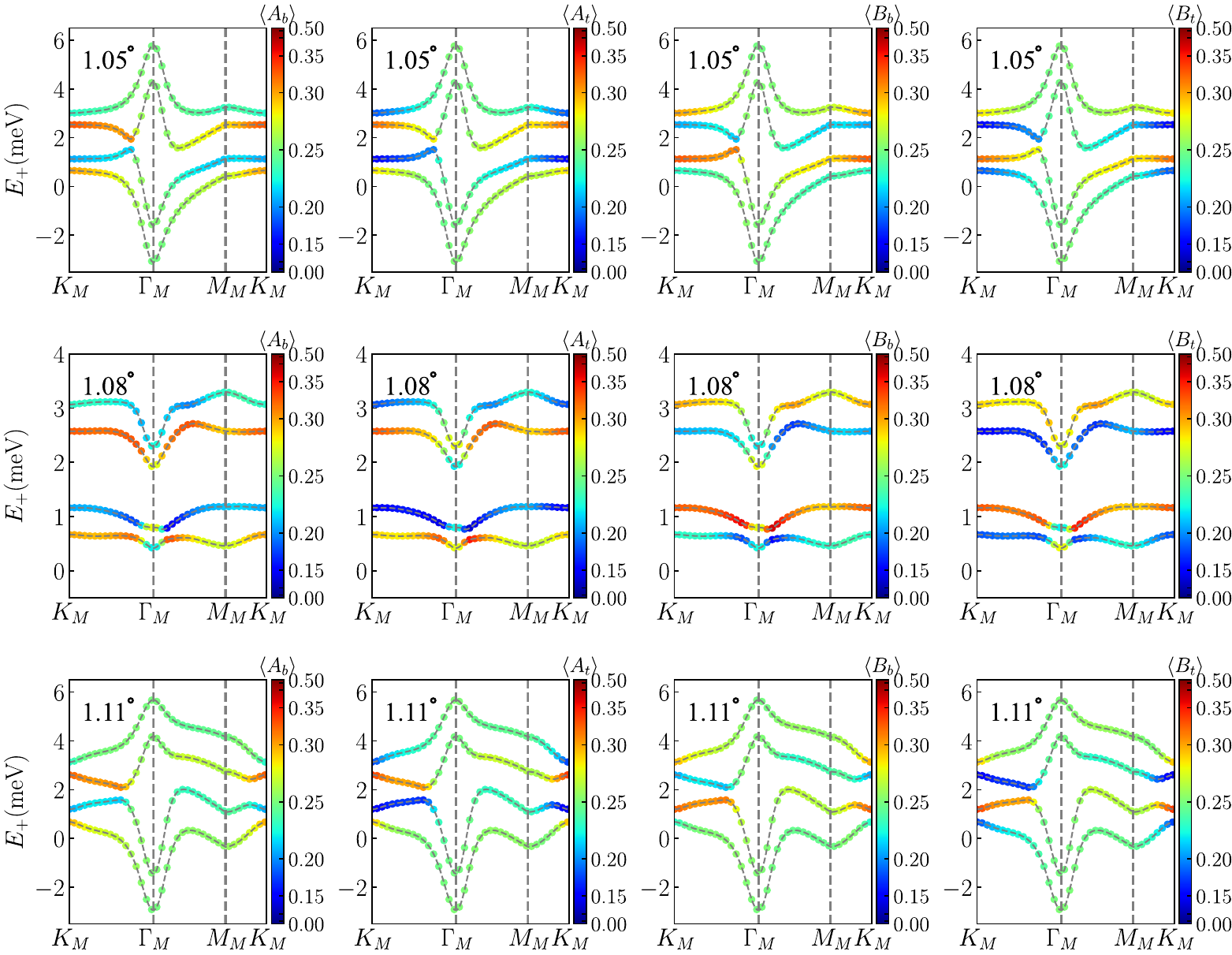}
	\caption{Layer-sublattice distribution for $\lambda_I=\lambda_R=3$meV. The mini bands in the $+K$ valley are plotted. $A_b$, $A_t$, $B_b$, and $B_t$ denote bottom layer sublattice $A$, top layer sublattice $A$, bottom layer sublattice $B$, and top layer sublattice $B$, respectively. Top row: $\theta=1.05^{\circ}$. Middle row: $\theta=1.08^{\circ}$. Bottom row: $\theta=1.11^{\circ}$}
	\label{Fig:LS_weights_I3_R3}
\end{figure}

It is also important to understand how SOC and twist angle influence the composition of wavefunction. Since the spin structure is plotted in Fig. 2 of the main text, we only show the layer-sublattice distribution in Figs.~\ref{Fig:LS_weights_I1_R1} and \ref{Fig:LS_weights_I3_R3}. At $1.08^\circ$, the sublattice polarization takes place for a wide range of $k$ points in each mini band. For $\theta=1.05^\circ$ and $\theta=1.11^\circ$, the sublattice polarization only becomes significant near the $K_M$ point. The SOC strengths also are important. The sublattice polarization effect in the $\lambda_I=\lambda_R=3$meV case is more pronounced than the $\lambda_I=\lambda_R=1$meV case. There are small differences between the top and bottom layers for all the cases, but those effects are not significant. The sublattice structures provide an explanation of why the interband superconductivity happens near the magic angle, in addition to the nesting between nearly flat bands. For $\theta>1.08^\circ$, the spin textures of the mini bands disfavor the interband intersublattice Ising superconductivity. Such an effect from spin texture is absent for $\theta<1.08^\circ$.

\section{Finite size effect}

In all the results presented in this work, we use a $24\times 24$ momentum grid. We obtain similar results using different sizes of momentum grids (e.g.,  $12\times 12$ and $36\times 36$) and/or different choices of first moir\'e Brillouin zone. The finite-size effects of the valley polarization and intervalley coherent order are negligible due to the large $T_c\sim 40$K. Meanwhile, the finite-size effect is most significant for the weak superconductivity ($T_c<50$mK), which is not our main focus. The overall qualitative trend is unaffected using different momentum grids in the linearized gap equation for superconductivity.


\begin{thebibliography}{121}
\bibitem [{IVC()}]{IVC}%
		\BibitemOpen
		\href@noop {} {}\bibinfo {note}
		{$\mathcal{V}^{(+,+),(bb'bb')}_{-\vex{k}',\vex{k}',\vex{k},-\vex{k}}$ is not
			manifestly gauge invariant, making numerical calculations challenging. Thus,
			we use the absolute value of
			$\mathcal{V}^{(+,+),(bb'bb')}_{-\vex{k}',\vex{k}',\vex{k},-\vex{k}}$, which
			likely overestimates the $T_{c,\alpha=x,y}$.}\BibitemShut {Stop}%
\bibitem [{Cou()}]{Coulomb}%
		\BibitemOpen
		\href@noop {} {}\bibinfo {note} {The Coulomb suppression is not considered in
			our calculations. Owing to the large density of states in the system, we
			expect that the Coulomb suppression is not significant (because of strong
			intraband screening by the carriers) and the qualitative conclusion remains
			unchanged.}\BibitemShut {Stop}%
\end{thebibliography}
\end{document}